\documentclass[
 groupedaddress,
 amsmath,
 amssymb,
 aps,
 pra,
 floatfix,
 twocolumn,
 longbibliography
]{revtex4-2}

\newcommand{\bra}[1]{\langle #1\rvert}
\newcommand{\ket}[1]{\lvert #1\rangle}

\newcommand{\op}[2]{\ket{#1} \bra{#2}}

\newcommand{\rpd}[1]{\partial_{#1}}
\newcommand{\rpdu}[1]{\partial^{#1}}
\newcommand{\rcd}[1]{\nabla_{#1}}
\newcommand{\rcdu}[1]{\nabla^{#1}}

\newcommand{\normalorder}[1]{:\!#1\!:}

\usepackage{slashed}
\usepackage{graphicx}
\usepackage{dcolumn}
\usepackage{bm}
\usepackage{hyperref}
\usepackage{tikz}
\usetikzlibrary{calc}

\usepackage{orcidlink}

\usepackage{amsmath}
\usepackage{upgreek}
\usepackage{physics}
\usepackage{placeins}
\usepackage{comment}
\usepackage{mathrsfs}
\usepackage{mathtools}
\usepackage{enumitem}

\begin{document}


\title{Non-Relativistic Quantum Electrodynamics of Atoms in a Rotating Ring Cavity}

\author{Jarrod T. Reilly\orcidlink{0000-0001-5410-089X}}
\affiliation{JILA and Department of Physics, University of Colorado, 440 UCB, Boulder, CO 80309, USA}
\author{Murray J. Holland\orcidlink{0000-0002-3778-1352}}
\affiliation{JILA and Department of Physics, University of Colorado, 440 UCB, Boulder, CO 80309, USA}

\date{\today}


\begin{abstract}
In this paper, we derive a quantum optics model in a non-inertial rotating ring cavity from first principles. 
We begin with the Dirac equation in curved spacetime, add minimal coupling to the electromagnetic field, and then find the Dirac Hamiltonian for the generalized Born metric. 
We then formally take the non-relativistic limit by way of Foldy-Wouthuysen transformations and project onto a fermionic Fock space for the atoms' electrons, protons, and neutrons. 
Focusing on a protium atom, we next move from a minimal coupling gauge to a multipole expansion gauge by taking a Power-Zienau-Woolley transformation under the dipole and long-wavelength approximations. 
Here, we find additional terms from the rotation of the system including a rotation-induced hyperfine shift of the atomic transition which could possibly be observed experimentally even for small rotation rates.
Making the electric dipole, two-level, and rotating-wave approximations, we arrive at a Jaynes-Cummings-like Hamiltonian which also accounts for rotational effects, such as the rotation-induced hyperfine shift and the Sagnac shift for the cavity's counterpropagating modes. 
\end{abstract}

{
\let\clearpage\relax
\maketitle
}

\section{Introduction}
Optical ring resonators have become a standard device used for exploring light-matter interactions~\cite{CohenTannoudji2}.
For example, they are utilized in experiments with a myriad of research objectives, such as quantum optics~\cite{Lee2,Ostermann,Chen,Zhou,Cline,Schafer2}, quantum metrology~\cite{Butt,Li2,Clark,Kresic}, quantum simulations~\cite{Mivehvar}, quantum information~\cite{Bao,Cox,Li3}, and biosensing~\cite{Ksendzov,Fard,Flueckiger}.
In particular, optical ring cavities play a central role in sensing rotations by way of active and passive optical gyroscopes~\cite{Chow,Faucheux,Lefevre} and Sagnac interferometers~\cite{Gauguet,Sun,Eberle}.
Here, the Sagnac effect~\cite{Post}, or more generally the Sagnac-Wang-Fizeau effect~\cite{Wang,Ori}, from the finite speed of light causes a rotating observer to see counterpropagating modes of the ring cavity acquire equal but opposite frequency shifts, and so the rotation rate can be read out from the interference pattern between the two modes.
State-of-the-art ring laser gyroscopes can sense remarkably small rotation rates and are one of the fundamental elements in modern inertial navigation systems on airplanes, satellites, and spacecraft~\cite{Vimal}. 
However, models of light-matter interactions in rotating ring cavities are typically restricted to a classical treatment of the electromagnetic (E\&M) fields, and so work on rotating quantum optics with a fully quantum description of the cavity field has been limited~\cite{Li,Shi,Zhu}.

A natural way to introduce inertial effects into a quantum optics model is to begin with a covariant quantum field theory description in which a non-inertial observer sees a modified spacetime.
Here, these effects can modify the spectral lines of atoms~\cite{Parker2,Parker3,Parker4}, introduce additional entanglement~\cite{Peres,Fuentes,Alsing,Bradler,Montero,Downes} and decoherence~\cite{Unruh,Wang2,Nesterov,Schwartz2,Jakubec}, generate spin currents~\cite{Matsuo,Matsuo2,Matsuo3,Funato}, alter the effective optical response in two-dimensional materials~\cite{Caneda}, and break typically assumed approximations in multipole interactions~\cite{Lopp}. 
We note that to introduce curved spacetime effects into quantum mechanics models, it is essential to begin with a quantum field theory model, introduce the curved spacetime, and then take the non-relativistic limit, as curved spacetime effects cannot be rigorously introduced at the level of quantum mechanics~\cite{Falcone,Falcone2}.
While quantum field theory descriptions of quantum optics models in Minkowski spacetime have garnered some recent interest~\cite{Lopp,Asano,Kattan}, covariant quantum field theory approaches to deriving quantum optics models in curved spacetimes have so far been limited~\cite{Burgess,Sorge}.

\begin{figure}
    \centerline{\includegraphics[width=\linewidth]{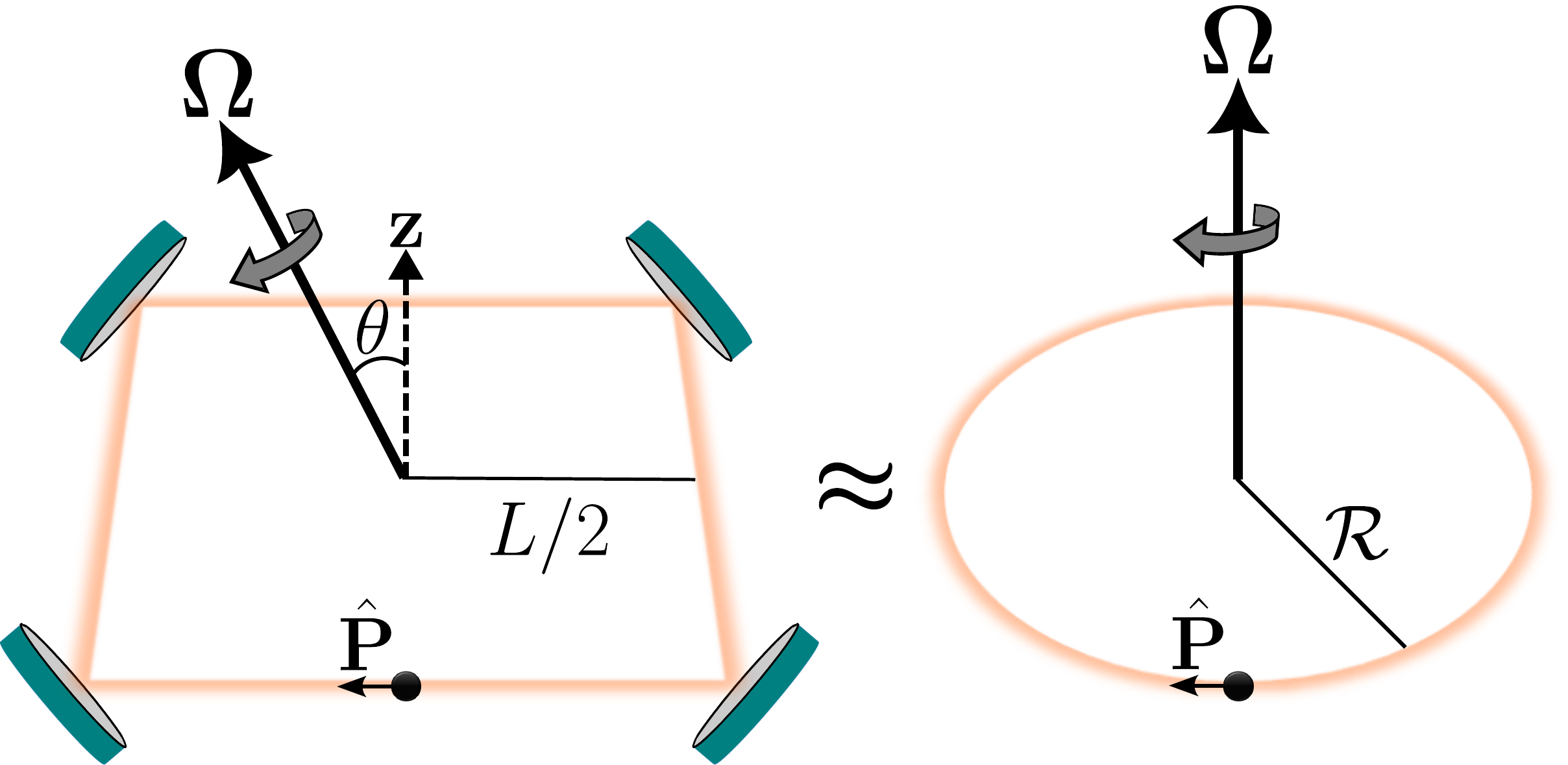}}
    \caption{Schematic diagram of a non-relativistic atom inside a ring cavity rotating with uniform angular velocity $\mathbf{\Omega}$.
    In our final Hamiltonian in Sec.~\ref{Sec:JaynesCummingsHydrogen}, we approximate the square cavity as a 1D ring and approximate $\mathbf{\Omega} \approx \tilde{\Omega} \mathbf{z} = \Omega \cos \theta \mathbf{z}$ so that we can use separation of variables in cylindrical coordinates (see Appendix~\ref{Appendix:EMquantization}).}
    \label{Schematic}
\end{figure}
In this paper, we derive a modified Jaynes-Cummings model for hydrogen atoms in a rotating ring cavity, depicted schematically in Fig.~\ref{Schematic}, from first principles. 
We take a different approach from Refs.~\cite{Burgess,Sorge}; we begin with a spinor field in curved spacetime coupled to the E\&M field for the atom's proton and electron rather than a scalar field with the two-level and long-wavelength approximations already made. 
We then formally take the non-relativistic limit of the Dirac Hamiltonian of the atom's particles by way of a Foldy-Wouthuysen (FW) transformation~\cite{Foldy,Bjorken,Silenko,Buhl,Goncalves,Case,Silenko2,Silenko3,Silenko4} which can accurately obtain relativistic corrections. 
We obtain a minimally coupled Hamiltonian for the hydrogen atom which we then take a Power-Zienau-Woolley (PZW) transformation~\cite{Power,Power2,Power3,Power4,Woolley,Woolley2,Babiker,Babiker2,CohenTannoudji,CohenTannoudji2,Andrews,Steck} of to obtain a multipole interaction Hamiltonian. 
Lastly, we make the typical electric dipole, two-mode (for clockwise and counterclockwise modes), two-level, and rotating-wave approximations to obtain a modified Jaynes-Cummings model Eq.~\eqref{H_final}.
From our general relativistic approach, we obtain two additional shifts at lowest-order: a rotation-induced hyperfine shift Eq.~\eqref{HyperfineShift} and a rotation-induced R\"ontgen interaction Eq.~\eqref{RotationRontgen}. 
These shifts, especially the former, may have an observable impact on measurements in optical lattice atomic clocks~\cite{Inguscio,Derevianko,Ludlow,Hollberg,Fortier,Bothwell}, superradiant lasers~\cite{Meiser,Norcia,Reilly}, and inertial sensors~\cite{Chow,Faucheux,Skulte,Reilly2}.

The structure of the paper is as follows.
In Sec.~\ref{Sec:DiracEqCurvedST}, we provide a brief introduction of the Dirac equation in curved spacetime minimally coupled to the E\&M field and apply the formalism to the generalized Born metric to arrive at a Dirac Hamiltonian Eq.~\eqref{H_D}.
In Sec.~\ref{Sec:FieldHamiltonian}, we derive the Hamiltonian of the free E\&M field in the rotating frame using the Arnowitt-Deser-Misner (ADM) formalism of general relativity.
We then take the non-relativistic limit of the Dirac Hamiltonian Eq.~\eqref{H_D} through a FW transformation in Sec.~\ref{Sec:NonRelLimit}.
Finally, we build a hydrogen atom from this non-relativistic Dirac field approach and take a PZW transformation to arrive at a modified Jaynes-Cummings model Eq.~\eqref{H_final} in Sec.~\ref{Sec:JaynesCummingsHydrogen}.
In Sec.~\ref{Sec:Conclusion}, we conclude with an outlook on generalizing our calculation to alkali(-like) and alkaline-earth(-like) atoms, as well as extending our methodology to other relevant spacetimes.

\section{Dirac Equation in Rotating Coordinates} \label{Sec:DiracEqCurvedST}
This section is dedicated to deriving the relativistic Dirac Hamiltonian of a spinor field for a rotating observer, Eq.~\eqref{H_D}. 
Since this is a derivation not commonly found in the literature of quantum optics, we will first slowly build up the Dirac equation in curved spacetime, model its interaction with the E\&M field through minimal coupling, and finally specify the metric tensor as the generalized Born metric, Eq.~\eqref{g_GB}, and calculate the resulting curved spacetime Dirac Hamiltonian. 

\subsection{Spin Connection}
We begin with the Dirac Lagrangian (density) in flat spacetime coming from a global $\mathrm{U} (1)$ symmetry of a spinor field $\Psi$~\cite{Schwartz,Tong_QFT},
\begin{equation}
    \mathcal{L}_{\mathrm{Df}} = \bar{\Psi} \left( i \slashed{\partial} - m \right) \Psi,
\end{equation}
where $\bar{\Psi} \equiv \Psi^{\dagger} \gamma^0$ is the Dirac adjoint, $\slashed{V} \equiv \gamma^{\mu} V_{\mu}$ is the Feynman slash notation, and we adopt natural units ($\hbar = c = 1$) until Sec.~\ref{Sec:JaynesCummingsHydrogen}. 
Here, $\rpd{\mu} = \partial/(\partial x^{\mu})$, $m$ is the rest mass, and $\gamma^{\mu}$ are the matrix representations of the Clifford algebra, $\{ \gamma^{\mu}, \gamma^{\nu} \} = 2 \eta^{\mu \nu}$ with the Minkowski metric $\eta_{\mu \nu} = \mathrm{diag} (1, -1, -1, -1) = \eta^{\mu \nu}$~\footnote{Note that we use a $(+,-,-,-)$ metric signature throughout the article.} (there is a implicit $4$-dimensional identity matrix on the right-hand-side of the anti-commutation relations). 
In particular, we are interested in the electron/positron field and the proton/anti-proton field, which we here take to be a point particle.
Thus, we use the Dirac-Pauli representation in which~\cite{Collas} $\gamma^0 = \beta$ and $\gamma^k = \beta \alpha^k$ with the Dirac matrices
\begin{equation} \label{AlphaBeta}
    \beta = 
\begin{pmatrix}
    \mathbb{I} & 0 \\ 0 & - \mathbb{I}
\end{pmatrix}, 
    \quad \alpha^k = 
\begin{pmatrix}
    0 & \sigma^k \\ \sigma^k & 0
\end{pmatrix},
\end{equation}
where $\mathbb{I} = \mathrm{diag}(1,1)$, $\sigma^k$ are the usual Pauli matrices, and we use lowercase Latin indices for purely spatial indices. 
The Euler-Lagrange equation with respect to $\bar{\Psi}$ then gives the Dirac equation,
\begin{equation}
    (i \slashed{\partial} - m) \Psi = 0.
\end{equation}

To upgrade the Dirac Lagrangian to curved spacetime, we require the spinor covariant derivative~\cite{Collas,Parker}, 
\begin{equation}
    \rpd{\mu} \longrightarrow \mathcal{D}_{\mu} \equiv \rpd{\mu} + \Gamma_{\mu},
\end{equation}
where the spinor affine connection is given by
\begin{equation}
    \Gamma_{\mu} = \frac{1}{4} \omega_{A B \mu} \gamma^A \gamma^B,
\end{equation}
with the spin connection coefficients~\footnote{The spin connection coefficients are usually defined in the form $\omega^A_{\phantom{A} B \mu}$, which are related to Eq.~\eqref{SpinConnection} through $\omega^A_{\phantom{A} B \mu} = \eta^{A C} \omega_{C B \mu}$.},
\begin{equation} \label{SpinConnection}
    \omega_{A B \mu} = g_{\alpha \beta} e_A^{\phantom{A} \beta} \rcd{\mu} e_B^{\phantom{B} \alpha} = \eta_{A C} e^C_{\phantom{C} \alpha} \rcd{\mu} e_B^{\phantom{B} \alpha}.
\end{equation}
Here, the spacetime-dependent tetrad (vielbein) vector fields $e_A = e_A^{\phantom{A} \mu} \rpd{\mu}$ and covector fields $e^A = e^A_{\phantom{A} \mu} dx^{\mu}$ can be thought of as the gravitational field~\cite{Collas,Chirstodoulou} through the relations between the metric tensor $g_{\mu \nu}$ and Minkowski metric~\cite{Collas,Misner,Weinberg,Carroll},
\begin{equation} \label{Tetrads}
    \begin{aligned}
g_{\mu \nu} (x) &= e^A_{\phantom{A} \mu} (x) e^B_{\phantom{B} \nu} (x) \eta_{AB}, \\ 
\eta_{AB} &= e_A^{\phantom{A} \mu} (x) e_B^{\phantom{B} \nu} (x) g_{\mu \nu} (x),
    \end{aligned}
\end{equation}
where we reserve capital Latin letters for the local ``flat'' spacetime indices (and so $A,B \in \{ 0, 1, 2, 3 \}$); see Refs.~\cite{Collas,Misner,Weinberg} for more details. 
Moreover, we have introduced the spacetime covariant derivative on vectors and one-forms~\cite{Misner,Weinberg,Carroll}
\begin{equation}
    \begin{aligned}
\rcd{\mu} V^{\nu} = \rpd{\mu} V^{\nu} + \Gamma_{\mu \lambda}^{\nu} V^{\lambda}, \\
\rcd{\mu} O_{\nu} = \rpd{\mu} O_{\nu} - \Gamma_{\mu \nu}^{\lambda} O_{\lambda},
    \end{aligned}
\end{equation}
with the Christoffel symbols from the Levi-Civita connection~\cite{Misner,Weinberg,Carroll},
\begin{equation}
    \Gamma_{\mu \nu}^{\lambda} = \frac{1}{2} g^{\lambda \rho} \left( \rpd{\mu} g_{\nu \rho} + \rpd{\nu} g_{\rho \mu} - \rpd{\rho} g_{\mu \nu} \right),
\end{equation}
which are symmetric in its lower indices for torsion-free metrics~\cite{Carroll}, $\Gamma_{\mu \nu}^{\lambda} = \Gamma_{\nu \mu}^{\lambda} = \Gamma_{(\mu \nu)}^{\lambda}$.
In addition to the spin connection, we must use the spacetime-dependent matrices~\cite{Collas,Birrell},
\begin{equation} \label{gammaCurved}
    \underline{\gamma}^{\mu} (x) = e_A^{\phantom{A} \mu} (x) \gamma^A,
\end{equation}
which now satisfy $\{ \underline{\gamma}^{\mu} (x), \underline{\gamma}^{\nu} (x) \} = 2 g^{\mu \nu} (x)$.
By upgrading $\rpd{\mu} \longrightarrow \mathcal{D}_{\mu}$ and $\gamma^{\mu} \longrightarrow \underline{\gamma}^{\mu}$, we obtain the Dirac Lagrangian in curved spacetime,
\begin{equation}
    \mathcal{L}_{\mathrm{D}} = \sqrt{- g} \bar{\Psi} \left( i \underline{\slashed{\mathcal{D}}} - m \right) \Psi,
\end{equation}
where we have introduced the spacetime-dependent slash notation $\underline{\slashed{V}} \equiv \underline{\gamma}^{\mu} (x) V_{\mu}$ and included the factor with the metric tensor's determinant $g$ to make the action integral, in the path integral approach to quantum field theory~\cite{Schwartz,Zee}, invariant under diffeomorphisms~\cite{Misner,Carroll}.
The Dirac equation in curved spacetime then becomes
\begin{equation}
    \left( i \underline{\slashed{\mathcal{D}}} - m \right) \Psi = 0.
\end{equation}
It is convenient to rewrite this in terms of the Fock-Ivanenko coefficients $\Gamma_K = e_K^{\phantom{K} \mu} \Gamma_{\mu}$~\cite{Collas},
\begin{equation}
    \left[ i \gamma^K \left( e_K + \Gamma_K \right) - m \right] \Psi = 0,
\end{equation}
where there's an implicit identity matrix on the $e_K$ term. 
In this, we can directly calculate the Fock-Ivanenko coefficients using~\cite{Collas}
\begin{equation} \label{FockIvan}
    \Gamma_K = \frac{1}{4} \Gamma_{A B K} \gamma^A \gamma^B,
\end{equation}
with the Ricci rotation coefficients~\cite{Collas},
\begin{equation} \label{RicciRot}
    \Gamma_{A B K} = \omega_{AB \mu} e_K^{\phantom{K} \mu} = - \frac{1}{2} \left( C_{ABK} + C_{BKA} - C_{KAB} \right),
\end{equation}
defined via the object of anholonomicity which encodes the non-commutativity of non-coordinate tetrads~\cite{Collas,Misner,Corum},
\begin{equation} \label{Anholon}
    C_{A B K} = \eta_{AD} C^D_{\phantom{D} B K} = \eta_{AD} \left( \rpd{\nu} e^D_{\phantom{D} \mu} - \rpd{\mu} e^D_{\phantom{D} \nu} \right) e_B^{\phantom{B} \mu} e_K^{\phantom{K} \nu},
\end{equation}
which are anti-symmetric in their first two and last two indices, respectively,
\begin{equation}
    \begin{aligned}
\Gamma_{ABK} &= - \Gamma_{BAK} = \Gamma_{[AB]K}, \\ 
C_{ABK} &= - C_{AKB} = C_{A[BK]}.
    \end{aligned}
\end{equation}
Note that the sign conventions for Eqs.~\eqref{FockIvan}--\eqref{Anholon} vary throughout the literature; we adopt the convention of Ref.~\cite{Collas}.

\subsection{Minimal Coupling to Electromagnetic Field}
To couple the Dirac field to the E\&M field, we introduce the Maxwell field (i.e., the four-vector potential) $A_{\mu} = (\varphi, - \mathbf{A})$ which gives the anti-symmetric Faraday tensor,
\begin{equation}
    F_{\mu \nu} = \rpd{\mu} A_{\nu} - \rpd{\nu} A_{\mu} = - F_{\nu \mu}.
\end{equation}
The Lagrangian for the E\&M field in flat spacetime is given by an abelian Yang-Mills theory~\cite{Schwartz,Tong_QFT,Tong_GT},
\begin{equation} \label{L_F}
    \mathcal{L}_{\mathrm{F}} = - \frac{1}{4} F^{\mu \nu} F_{\mu \nu},
\end{equation}
which we take to be in the Coulomb gauge $\bm{\nabla} \bm{\cdot} \mathbf{A} = 0$, where boldface signifies a spatial vector.
We can then couple this $\mathrm{U} (1)$ gauge field to the Dirac field of charge $q$ by enforcing a local $\mathrm{U} (1)$ symmetry on $\Psi$~\cite{Tong_QFT,Tong_GT,Parker} which, for a minimal coupled model, amounts to replacing partial derivative by the gauge covariant derivative~\cite{Schwartz,Tong_QFT},
\begin{equation}
    \rpd{\mu} \longrightarrow D_{\mu} \equiv \rpd{\mu} + i q A_{\mu}.
\end{equation}
This leads to the Dirac Lagrangian interacting with the E\&M field~\cite{Tong_QFT,Schwartz},
\begin{equation}
    \mathcal{L}_{\mathrm{If}} = \bar{\Psi} \left( i \slashed{D} - m \right) \Psi - \frac{1}{4} F^{\mu \nu} F_{\mu \nu},
\end{equation}
which then gives important results from relativistic quantum electrodynamics such as the Lamb shift~\cite{Lamb,Bethe2,Bethe,Schwartz} and the anomalous magnetic dipole moment~\cite{Schwinger,Feynman,Fan,Schwartz}. 

Similar to the previous subsection, in curved spacetime we must now upgrade the spinor covariant derivative to a total covariant derivative that includes the gauge covariant derivative~\cite{Collas},
\begin{equation}
    \mathcal{D}_{\mu} \longrightarrow \mathfrak{D}_{\mu} \equiv D_{\mu} + \Gamma_{\mu} = \rpd{\mu} + \Gamma_{\mu} + i q A_{\mu}.
\end{equation}
With this, the interacting Lagrangian in curved spacetime is given by
\begin{equation}
    \mathcal{L}_{\mathrm{I}} = \sqrt{- g} \left[ \bar{\Psi} \left( i \underline{\slashed{\mathfrak{D}}} - m \right) \Psi - \frac{1}{4} F^{\mu \nu} F_{\mu \nu} \right],
\end{equation}
which gives the curved spacetime Dirac equation from the Euler-Lagrange equation with respect to $\bar{\Psi}$,
\begin{equation}
    \left( i \underline{\slashed{\mathfrak{D}}} - m \right) \Psi = 0. 
\end{equation}
Defining $A_K = e_K^{\phantom{K} \mu} A_{\mu}$, we can rewrite this in a particularly convenient way in terms of Fock-Ivanenko coefficients~\cite{Collas},
\begin{equation} \label{DiracEqFI}
    \left[ i \gamma^K \left( e_K + \Gamma_K + i q A_K \right) - m \right] \Psi = 0.
\end{equation}

\subsection{Generalized Born Metric} \label{Sec:GenBornMetric}
The derivation of the tetrads for a non-inertial observer traveling through a flat spacetime is covered in detail in Refs.~\cite{Misner,Ni,Hehl}, and so we only give a brief overview here. 
We begin with the metric for a Langevin observer~\cite{Langevin} rotating with some uniform angular velocity vector $\mathbf{\Omega}$ with respect to an inertial frame. 
We can transform from Minkowski coordinates $(T, \mathbf{r}')$ to the local rotating coordinates $(t,\mathbf{r})$ using the coordinate chart~\cite{Misner,Ni,Hehl}
\begin{equation}
    dt = dT, \quad d \mathbf{r} = d \mathbf{r}' - \left( \mathbf{\Omega} \cross \mathbf{r} \right) dt.
\end{equation}
Thus, the metric tensor becomes
\begin{equation} \label{g_GB}
    \begin{aligned}
ds^2 &= g_{\mu \nu} dx^{\mu} dx^{\nu} \\
&= \left[ 1 - \left( \mathbf{\Omega} \cross \mathbf{r} \right)^2 \right] dt^2 - 2 \left( \mathbf{\Omega} \cross \mathbf{r} \right) \bm{\cdot} d \mathbf{r} dt - d \mathbf{r}^2,
    \end{aligned}
\end{equation}
which we will refer to as the generalized Born metric as it extends the typical Born metric Eq.~\eqref{g_B} to an arbitrary rotation axis. 
Since $g_{\mu \nu}$ is independent of $t$, there is a Killing vector $K^{\mu} = (\rpd{t})^{\mu}$~\cite{Misner,Weinberg,Carroll} whose magnitude is given by (temporarily inserting factors of $c$)
\begin{equation} \label{KillingMag}
    K_{\mu} K^{\mu} = g_{tt} = \left[ c^2 + \left( \mathbf{\Omega} \bm{\cdot} \mathbf{r} \right)^2 - \Omega^2 \left( \mathbf{r} \bm{\cdot} \mathbf{r} \right) \right].
\end{equation}
We therefore always assume that $\Omega \abs{\mathbf{r}} < c$ such that $K^{\mu}$ is timelike (i.e., our observer is within the light cylinder).

Noting that the differential line element can be written in terms of tetrad covectors as~\cite{Misner,Hehl}
\begin{equation}
    ds^2 = e^{\hat{0}} \otimes e^{\hat{0}} - \sum_{j = 1}^3 e^{\hat{j}} \otimes e^{\hat{j}},
\end{equation}
we have
\begin{equation}
    e^{\hat{0}} = dt, \quad e^{\hat{j}} = dx^j + \left[ \mathbf{\Omega} \cross \mathbf{r} \right]^j dt,
\end{equation}
where the hat here is used to represent the observer's local coordinates~\cite{Hehl,Carroll} and $[\bm{\cdot}]^j$ stands for the $j$th component of the vector. 
Using Eq.~\eqref{Tetrads}, we then find the tetrad vectors
\begin{equation}
    e_{\hat{0}} = \rpd{t} - \left[ \mathbf{\Omega} \cross \mathbf{r} \right]^j \rpd{j}, \quad e_{\hat{j}} = \rpd{j}. 
\end{equation}
From this, we obtain the Ricci rotation coefficients,
\begin{equation}
    \Gamma_{\hat{i} \hat{j} \hat{0}} = \varepsilon_{ijk} \Omega^k, \quad \Gamma_{\hat{0} \hat{0} \hat{0}} = \Gamma_{\hat{0} \hat{j} \hat{0}} = \Gamma_{\hat{\mu} \hat{\nu} \hat{j}} = 0,
\end{equation}
where $\varepsilon_{ijk}$ is the typical Levi-Civita pseudotensor on the spatial coordinates. 
Finally, we arrive at the Fock-Ivanenko coefficients,
\begin{equation}
    \Gamma_{\hat{0}} = \frac{1}{4} \varepsilon_{ijk} \Omega^k \gamma^i \gamma^j = \frac{1}{4} \mathbf{\Omega} \bm{\cdot} \left[ \beta \bm{\alpha} \cross \beta \bm{\alpha} \right], \quad \Gamma_{\hat{j}} = 0,
\end{equation}
in agreement with Refs.~\cite{Hehl,Matsuo,Matsuo2}. 
Expanding and inserting the identity~\cite{Sakurai} $\bm{\sigma} \cross \bm{\sigma} = 2 i \bm{\sigma}$, we get
\begin{equation}
    \begin{aligned}
\beta \bm{\alpha} \cross \beta \bm{\alpha} &= - 
    \begin{pmatrix}
\bm{\sigma} \cross \bm{\sigma} & 0 \\
0 & \bm{\sigma} \cross \bm{\sigma}
    \end{pmatrix}
&= - 2 i
    \begin{pmatrix}
\bm{\sigma} & 0 \\
0 & \bm{\sigma}
    \end{pmatrix},
    \end{aligned}
\end{equation}
such that, dropping the identity matrix,
\begin{equation} \label{GB_SpinConnection}
    \Gamma_{\hat{0}} = - i \mathbf{\Omega} \bm{\cdot} \mathbf{S},
\end{equation}
where we have defined the spin operator as (in SI units) $\mathbf{S} = \hbar \bm{\sigma} / 2$.

Collecting terms and inserting them into Eq.~\eqref{DiracEqFI}, we obtain 
\begin{equation}
    \begin{aligned}
i \beta \left[ \rpd{t} - i \left( \mathbf{\Omega} \cross \mathbf{r} \right) \bm{\cdot} \left( \mathbf{p} - q \mathbf{A} \right) - i \mathbf{\Omega} \bm{\cdot} \mathbf{S} + i q \varphi \right] \Psi = \\
- i \beta \bm{\alpha} \bm{\cdot} \left[ i \mathbf{p} - i q \mathbf{A} \right] \Psi + m \Psi,
    \end{aligned}
\end{equation}
where we have defined the usual momentum operator $\mathbf{p} = - i \hbar \bm{\nabla}$ and used $A_{\hat{0}} = A_0 + (\mathbf{\Omega} \cross \mathbf{r}) \bm{\cdot} \mathbf{A}$. 
Multiplying this from the left by $\beta$ and using $\beta^2 = \mathbb{I}_4$ with identity matrix $\mathbb{I}_4 = \mathrm{diag} (1,1,1,1)$, we find
\begin{equation}
    \begin{aligned}
\left( i \rpd{t} + \mathbf{\Omega} \bm{\cdot} \left[ \mathbf{r} \cross \left( \mathbf{p} - q \mathbf{A} \right) \right] + \mathbf{\Omega} \bm{\cdot} \mathbf{S} - q \varphi \right) \Psi = \\
\bm{\alpha} \bm{\cdot} \left( \mathbf{p} - q \mathbf{A} \right) \Psi + \beta m \Psi,
    \end{aligned}
\end{equation}
where we have used~\cite{Griffiths_EM} $(\mathbf{a} \cross \mathbf{b}) \bm{\cdot} \mathbf{c} = \mathbf{a} \bm{\cdot} (\mathbf{b} \cross \mathbf{c})$.
Identifying the angular momentum operator as $\mathbf{L} = \mathbf{r} \cross \mathbf{p}$ and the total angular momentum operator as $\mathbf{J} = \mathbf{L} + \mathbf{S}$, we find
\begin{equation}
    i \rpd{t} \Psi = \left[ \beta m + \bm{\alpha} \bm{\cdot} \left( \mathbf{p} - q \mathbf{A} \right) + q \varphi - \mathbf{\Omega} \bm{\cdot} \left( \mathbf{J} - q \mathbf{r} \cross \mathbf{A} \right) \right] \Psi.
\end{equation}
Temporarily reinserting SI units, we obtain the Dirac equation in Hamiltonian form
\begin{equation}
    i \hbar \rpd{t} \Psi = H_{\mathrm{D}} \Psi,
\end{equation}
with the Dirac Hamiltonian 
\begin{equation} \label{H_D}
    H_{\mathrm{D}} = \beta m c^2 + c \bm{\alpha} \bm{\cdot} \left( \mathbf{p} - q \mathbf{A} \right) + q \varphi - \mathbf{\Omega} \bm{\cdot} \left( \mathbf{J} - q \mathbf{r} \cross \mathbf{A} \right). 
\end{equation}
The Dirac Hamiltonian is Hermitian under the scalar product of $\Psi$ (with a metric-dependent measure)~\cite{Parker,Falcone,Falcone3} since the metric tensor is stationary~\cite{Wald,Parker2,Huang,Collas}, $\rpd{t} g_{\mu \nu} = 0$ and $g_{0j} \neq 0$ for some $j$~\cite{Misner,Carroll}. 

The Dirac Hamiltonian Eq.~\eqref{H_D} is the main result of this section. 
However, for future use, we consider a specific case of $\mathbf{\Omega} = (0, 0, \tilde{\Omega})^T$ [later, we will take $\tilde{\Omega}$ to be the component of $\mathbf{\Omega}$ perpendicular to the plane of the cavity]. 
Specifying this rotation axis and moving into cylindrical coordinates $(\rho,\phi,z)$ reduces $g_{\mu \nu}$ to the typical Born metric (again in natural units)~\cite{Born,Rindler,Muller,Soler},
\begin{equation} \label{g_B}
    \begin{aligned}
ds^2 &= g_{\mu \nu}^{\mathrm{B}} dx^{\mu} dx^{\nu} \\
&= (1 - \rho^2 \tilde{\Omega}^2) dt^2 - 2 \rho^2 \tilde{\Omega} dt d \phi - d \rho^2 - \rho^2 d \phi^2 - dz^2.
    \end{aligned}
\end{equation}
We will, for the most part, use the generalized Born metric Eq.~\eqref{g_GB} in our model, but we will simplify our discussion by using the reduced Born metric Eq.~\eqref{g_B} when we quantize the E\&M field inside the ring cavity in Secs.~\ref{Sec:EMsimplify} and~\ref{Sec:Approxs}, as well as the end of Appendix~\ref{Appendix:EMquantization}.

\section{Electromagnetic Field Hamiltonian} \label{Sec:FieldHamiltonian}
We now wish to find the E\&M field Hamiltonian in the Coulomb gauge, $\bm{\nabla} \bm{\cdot} \mathbf{A} = 0$. 
To do this, we use the Arnowitt-Deser-Misner (ADM) formalism of general relativity~\cite{Arnowitt,Arnowitt2,Jha,Corichi} which decomposes spacetime into a $(3+1)$-foliation of space and time.
This allows a natural Hamiltonian description of general relativity to be developed in which the theory describes time-varying fields on Cauchy hypersurfaces $\Sigma_t$.
To connect the Cauchy hypersurface $\Sigma_t$ to the next parallel hypersurface $\Sigma_{t + dt}$, one introduces the normal vector to the hypersurface $n^{\mu}$, the lapse function $N$, and the vector of shift functions $N^j$. 
The lapse function can be thought of as determining the proper time to go between $\Sigma_t$ and $\Sigma_{t + dt}$, $d\tau = N dt$, while the shift function relates how the spatial coordinates have changed between hypersurfaces~\cite{Jha,Barrau}.
They are given by $N = 1 / \sqrt{g^{00}}$ and $N^j = N^2 g^{0j}$, while the covariant normal vector is simply $n_{\mu} = (N,0,0,0)$.
Lastly, we can define the three-dimensional spatial metric $\gamma_{jk}$ on the Cauchy hypersurface that is induced by $g_{\mu \nu}$.
Putting this all together, we obtain the metric and inverse metric in the form of a block matrix,
\begin{equation}
    \begin{aligned}
g_{\mu \nu} &\dot{=} \left( 
\begin{array}{c|c}
    N^2 - N_j N^j & N_k \\ 
    \hline
    N_j & - \gamma_{jk} \\
\end{array} 
\right), \\ 
g^{\mu \nu} &\dot{=} \frac{1}{N^2} \left( 
\begin{array}{c|c}
    1 & N^k \\ 
    \hline
    N^j & N^j N^k - N^2 \gamma^{jk} \\
\end{array} 
\right),
    \end{aligned}
\end{equation}
where $N_j = \gamma_{jk} N^k$ and $\gamma_{jk}$ is defined to have a $(+,+,+)$ signature.
The spatial metric $\gamma_{jk}$ in turn defines a spatial covariant derivative $D_j$ over the Cauchy hypersurface such that $D_j \gamma_{kl} = D_j \gamma^{kl} = 0$ (see Ref.~\cite{Jha} for the explicit relationship between $D_j$ and $\rcd{\mu}$).
We further note that the metric determinant becomes $\sqrt{- g} = N \sqrt{\gamma}$. 
For the generalized Born metric Eq.~\eqref{g_GB}, we thus obtain (in natural units)
\begin{equation} \label{GenBorn_ADM}
    N = 1, \quad N^j = - \left[\mathbf{\Omega} \cross \mathbf{r}\right]^j, \quad \gamma_{jk} = \gamma^{jk} = \mathrm{diag}(1,1,1).
\end{equation}

We can now turn to the E\&M Lagrangian in curved spacetime,
\begin{equation}
    \mathcal{L}_{\mathrm{F}} = - \frac{\sqrt{- g}}{4} F^{\mu \nu} F_{\mu \nu}.
\end{equation}
Note that the field strength tensor expression is manifestly covariant even with partial derivatives,
\begin{equation}
    F_{\mu \nu} = \rcd{\mu} A_{\nu} - \rcd{\nu} A_{\mu} = \rpd{\mu} A_{\nu} - \rpd{\nu} A_{\mu},
\end{equation}
for torsion-free metrics~\cite{Carroll} $\Gamma_{\mu \nu}^{\lambda} = \Gamma_{\nu \mu}^{\lambda}$.
We can then find the conjugate momentum to the field,
\begin{equation} \label{ConjMom}
    \Pi^j = \frac{\partial \mathcal{L}_{\mathrm{F}}}{\partial (\rpd{0} A_j)} = N \sqrt{\gamma} F^{j0} = - \frac{\sqrt{\gamma}}{N} \gamma^{jk} \left[ F_{k 0} - N^{i} F_{i k} \right],
\end{equation}
where we note that $\Pi^0 = 0$~\cite{Tong_QFT} and the last term differs in sign to that in Ref.~\cite{Jha} due to the use of a different metric signature. 
We can then obtain the Hamiltonian (density) by taking a Legendre transformation,
\begin{equation}
    \begin{aligned}
\mathcal{H}_{\mathrm{F}} &= \Pi^{\mu} \rpd{0} A_{\mu} - \mathcal{L}_{\mathrm{F}} = \Pi^j \rpd{0} A_j - \mathcal{L}_{\mathrm{F}} \\
&= \frac{N}{2 \sqrt{\gamma}} \Pi^j \Pi_j + \frac{N \sqrt{\gamma}}{4} F^{j k} F_{j k} - N^j \Pi^k F_{jk} + \Pi^j \rpd{j} A_0.
    \end{aligned}
\end{equation}
Since the covariant derivative of a scalar field is just the partial derivative~\cite{Misner,Carroll,Weinberg}, we thus have $D_j A_0 = \rpd{j} A_0$.
Inserting this into the field Hamiltonian and integrating by parts, we thus obtain
\begin{equation} \label{H_F_ADM}
    \mathcal{H}_{\mathrm{F}} = \frac{N}{2 \sqrt{\gamma}} \Pi^j \Pi_j + \frac{N \sqrt{\gamma}}{4} F^{j k} F_{j k} - N^j \Pi^k F_{jk} - A_0 D_j \Pi^j.
\end{equation}

We can then introduce the electric and magnetic fields via~\cite{Jha}
\begin{equation} \label{EBdef}
    E^j \equiv E^{\mu} = n_{\nu} F^{\mu \nu}, \quad B^j \equiv B^{\mu} = n_{\nu} G^{\mu \nu},
\end{equation}
where we have defined the dual of the Faraday tensor via the Hodge star operator~\cite{Misner},
\begin{equation}
    G^{\mu \nu} = (\star F)^{\mu \nu} = \frac{1}{2} \varepsilon^{\mu \nu \alpha \beta} F_{\alpha \beta}, 
\end{equation}
with the four-dimensional Levi-Civita pseudotensor $\varepsilon^{\mu \nu \alpha \beta}$~\cite{Landau} and implicit musical isomorphisms~\cite{Lee} [e.g., $\star F \dot{=} \star (F^{\flat})$].
We thus have
\begin{equation}
    F^{j0} = E^j / N, \quad F_{jk} = - \varepsilon_{jkl} B^l,
\end{equation}
where we have used $\varepsilon^{jkl} = n_{\nu} \varepsilon^{\nu jkl} = N \varepsilon^{0jkl}$~\cite{Jha} as well as $\varepsilon_{ijk} \varepsilon^{mnk} = \delta_i^m \delta_j^n - \delta_i^n \delta_j^m$~\cite{Landau}.
Inserting these into the Hamiltonian Eq.~\eqref{H_F_ADM}, we obtain
\begin{equation}
    \mathcal{H}_{\mathrm{F}} = \frac{N \sqrt{\gamma}}{2} \left( E^2 + B^2 \right) + \sqrt{\gamma} \left( \varepsilon_{jkl} N^j E^k B^l - A_0 D_j E^j \right),
\end{equation}
where we have used $\varepsilon_{jkl} \varepsilon^{ikl} = 2 \delta_j^i$~\cite{Landau}, $D_j \sqrt{\gamma} = 0$, and defined $E^2 = \gamma_{jk} E^j E^k$ and $B^2 = \gamma_{jk} B^j B^k$.
Noting that the final term is simply a Lagrange multiplier $A_0$ that imposes Gauss's law for electric fields in curved spacetime~\cite{Thorne,Jha}, we can set this term to zero for the case of no free charges $D_j E^j = 0$.
Thus, we obtain
\begin{equation}
    \mathcal{H}_{\mathrm{F}} = \frac{N \sqrt{\gamma}}{2} \left( E^2 + B^2 \right) + \sqrt{\gamma} \mathbf{N} \bm{\cdot} (\mathbf{E} \cross \mathbf{B}).
\end{equation}
Finally, we can insert the specific form of the generalized Born metric Eq.~\eqref{GenBorn_ADM} which gives
\begin{equation}
    \mathcal{H}_{\mathrm{F}} = \frac{1}{2} \left( E^2 + B^2 \right) - \left( \mathbf{\Omega} \cross \mathbf{r} \right) \bm{\cdot} (\mathbf{E} \cross \mathbf{B}),
\end{equation}
where now we can identify $E^2 = \mathbf{E} \bm{\cdot} \mathbf{E}$ and $B^2 = \mathbf{B} \bm{\cdot} \mathbf{B}$ since space is flat over the hypersurface.
Then using the scalar triple product identity~\cite{Jackson}, $(\mathbf{a} \cross \mathbf{b}) \bm{\cdot} \mathbf{c} = \mathbf{a} \bm{\cdot} (\mathbf{b} \cross \mathbf{c})$, we can rewrite the final term as
\begin{equation}
    (\mathbf{\Omega} \cross \mathbf{r}) \bm{\cdot} (\mathbf{E} \cross \mathbf{B}) = \mathbf{\Omega} \bm{\cdot} [\mathbf{r} \cross (\mathbf{E} \cross \mathbf{B})].
\end{equation}
Thus, the Hamiltonian becomes
\begin{equation} \label{H_F_vec}
    \mathcal{H}_{\mathrm{F}} = \frac{1}{2} \left( E^2 + B^2 \right) - \mathbf{\Omega} \bm{\cdot} \bm{\ell},
\end{equation}
where we have identified the angular momentum density of the E\&M field as (in SI units)~\cite{Griffiths_EM}
\begin{equation} \label{ell}
    \bm{\ell} = \epsilon_0 \mathbf{r} \cross (\mathbf{E} \cross \mathbf{B}).
\end{equation}
Lastly, we can quantize the E\&M field and plug it into the field Hamiltonian, 
\begin{equation} \label{H_F}
    \hat{H}_{\mathrm{F}} = \int d^3 r \left[ \frac{1}{2} \left( \hat{E}^2 + \hat{B}^2 \right) - \mathbf{\Omega} \bm{\cdot} \hat{\bm{\ell}} \right],
\end{equation}
which we do in Appendix~\ref{Appendix:EMquantization} and discuss further in Sec.~\ref{Sec:JaynesCummingsHydrogen}.

\section{Non-Relativistic Limit} \label{Sec:NonRelLimit}
With the E\&M Hamiltonian for the rotating observer worked out, we now wish to take the non-relativistic limit of the minimally coupled Dirac Hamiltonian from Eq.~\eqref{H_D}.
To do this, we write the Hamiltonian as
\begin{equation}
    H_{\mathrm{D}} = \beta m + \mathcal{E} + \mathcal{O},
\end{equation}
with even and odd components~\cite{Foldy,Bjorken}
\begin{equation} \label{EvenOdd}
    \begin{aligned}
\mathcal{E} &= q \varphi - \mathbf{\Omega} \bm{\cdot} \mathbf{J} + q \mathbf{\Omega} \bm{\cdot} \left( \mathbf{r} \cross \mathbf{A} \right), \\ 
\mathcal{O} &= \bm{\alpha} \bm{\cdot} (\mathbf{p} - q \mathbf{A}),
    \end{aligned}
\end{equation}
which commutate and anti-commute with $\beta$, respectively~\cite{Bjorken}, $\beta \mathcal{E} = \mathcal{E} \beta$ and $\beta \mathcal{O} = - \mathcal{O} \beta$. 
It is thus clear that it is the odd terms that will couple the ``large'' and ``small'' components of the Dirac spinor, and so we can systematically decouple the two by block-diagonalizing the Hamiltonian through successive unitary transforms. 
This is accomplished using the Foldy-Wouthuysen (FW) transformation~\cite{Foldy,Bjorken,Silenko} $U_{\mathrm{FW}} = e^{i S}$ with the Hermitian operator
\begin{equation}
    S = -\frac{i \beta \mathcal{O}}{2 m}.
\end{equation}
Three successive FW transformations of the resulting odd components lead to a block-diagonal (even) Hamiltonian to order $\mathcal{O} (1 / m^2)$ for a Dirac spinor coupled to external E\&M fields~\cite{Foldy,Bjorken,Hehl,Matsuo,Matsuo2}.
This transformation method implicitly assumes that $\mathcal{E}$ and $\mathcal{O}$ are of order $\mathcal{O}(1 / m^n)$ for $n \geq 0$~\cite{Buhl}, and so we must take the small rotation and weak E\&M field limits.

Since Refs.~\cite{Foldy,Bjorken} consider the case of a Dirac particle coupled to an external field in detail, and Refs.~\cite{Hehl,Matsuo2} cover the rotating coordinate case, we simply state the results of the three successive FW transformations here.
More details of our specific FW transformation are found in Appendix~\ref{Appendix:FWtransform}.
We find that the equation of motion for the transformed states is given by
\begin{equation}
    i \rpd{t} \Psi' = \left[ H_{\mathrm{D}}' + \mathrm{hoc} \right] \Psi',
\end{equation}
where $H_{\mathrm{D}}'$ is the Hamiltonian to order $\mathcal{O} (1 / m^2)$~\cite{Foldy,Bjorken}
\begin{equation} \label{FWtransform}
    \begin{aligned}
H_{\mathrm{D}}' = & \beta \left( m + \frac{\mathcal{O}^2}{2 m} - \frac{\mathcal{O}^4}{8 m^3} \right) + \mathcal{E} \\
& - \frac{1}{8 m^2} \left[ \mathcal{O}, \left( \left[ \mathcal{O}, \mathcal{E} \right] + i \rpd{t} \mathcal{O} \right) \right],
    \end{aligned}
\end{equation}
while $\mathrm{hoc}$ represents higher-order corrections that are $\mathcal{O} (1 / m^n)$ for $n > 2$, and thus not necessarily block-diagonal.
Plugging in our $\mathcal{E}$ and $\mathcal{O}$ from Eq.~\eqref{EvenOdd} and defining the electric and magnetic fields from Eq.~\eqref{EBdef},
\begin{equation}
    \mathbf{E} = - \bm{\nabla} \varphi - \rpd{t} \mathbf{A} - \left( \mathbf{\Omega} \cross \mathbf{r} \right) \cross \mathbf{B}, \quad \mathbf{B} = \bm{\nabla} \cross \mathbf{A}, 
\end{equation}
we find
\begin{equation} \label{H'_D}
    \begin{aligned}
H_{\mathrm{D}}' = & \beta \left[ m + \frac{(\mathbf{p} - q \mathbf{A})^2}{2 m} - \frac{q}{m} \mathbf{S} \bm{\cdot} \mathbf{B} \right] + q \varphi \\
& - \mathbf{\Omega} \bm{\cdot} \left[ \mathbf{J} - q \left( \mathbf{r} \cross \mathbf{A} \right) \right] \\
& + \frac{q}{4 m^2} \mathbf{S} \bm{\cdot} \left[ \left( \mathbf{p} - q \mathbf{A} \right), \mathbf{E} \right]_{\cross} - \frac{i q}{8 m^2} \mathbf{p} \bm{\cdot} \mathbf{E} - \frac{\beta \mathbf{p}^4}{8 m^3},
    \end{aligned}
\end{equation}
where $[\mathbf{A}, \mathbf{B}]_{\cross} = \mathbf{A} \cross \mathbf{B} - \mathbf{B} \cross \mathbf{A}$ is the cross product commutator.
The first line represents the typical lowest-order terms~\cite{Foldy,Bjorken} while the second line represents a lowest-order correction due to the inertial effects from the rotation~\cite{Hehl}. 
Next, the third line has the spin-orbit coupling of the electron and the Darwin term~\cite{Foldy,Bjorken}, which are the typical leading-order relativistic corrections~\cite{Barut}, as well as a relativistic correction to the particle's momentum. 
The electric fields in these terms include the inertial corrections to the spin-orbit coupling and Darwin terms due to the rotation~\cite{Matsuo,Matsuo2}. 

With the Dirac Hamiltonian now block diagonal to order $\mathcal{O}(1 / m^2)$, we can now go from a relativistic field theory to a non-relativistic field theory. 
We assume the transformed Dirac spinor can be decomposed as~\cite{Goncalves}
$\Psi' = \binom{\psi}{\chi} e^{- i m t}$ such that,
\begin{equation}
    i \rpd{t}
\begin{pmatrix}
    \psi \\ \chi
\end{pmatrix} \approx \left[ H_{\mathrm{D}}' - m \right] 
\begin{pmatrix}
    \psi \\ \chi
\end{pmatrix}.
\end{equation}
Remembering that $\beta m$ in $H_{\mathrm{D}}'$ is positive for $\psi$ and negative for $\chi$ [see Eq.~\eqref{AlphaBeta}], we thus find that the rest-mass-energy cancels for particles $\psi$ but adds to $- 2 m$ for anti-particles $\chi$ such that, for non-relativistic energies $\abs{i \rpd{t} \psi} \ll \abs{m \psi}$ and $\abs{i \rpd{t} \chi} \ll \abs{m \chi}$, we can focus on the Pauli-Schr\"odinger equation for the particles,
\begin{equation}
    i \rpd{t} \psi = H_{\mathrm{NR}} \psi,
\end{equation}
with a non-relativistic Hamiltonian acting solely on the $\psi$-sector (positive-energy-sector)~\cite{Goncalves}
\begin{equation} \label{H_NR}
    H_{\mathrm{NR}} = H_0 + H_{\mathrm{r}}.
\end{equation}
Here, we have broken the Hamiltonian up into the leading-order term
\begin{equation} \label{H_0}
    H_0 = \frac{(\mathbf{p} - q \mathbf{A})^2}{2 m} - \frac{g q}{2 m} \mathbf{S} \bm{\cdot} \mathbf{B} + q \varphi - \mathbf{\Omega} \bm{\cdot} \left[ \mathbf{J} - q \left( \mathbf{r} \cross \mathbf{A} \right) \right],
\end{equation}
and the lowest-order relativistic correction,
\begin{equation} \label{H_r}
    \begin{aligned}
H_{\mathrm{r}} = & \frac{q}{4 m^2} \mathbf{S} \bm{\cdot} \left[ \left( \mathbf{p} - q \mathbf{A} \right), \mathbf{E} \right]_{\cross} - \frac{i q}{8 m^2} \mathbf{p} \bm{\cdot} \mathbf{E} - \frac{\mathbf{p}^4}{8 m^3}.
    \end{aligned}
\end{equation}
Here, $g$ is the spin $g$-factor which is $g = 2$ for a Dirac field without QED corrections. 

Lastly, we pass from the FW-reduced Dirac field theory to an effective non-relativistic particle description by projecting onto the positive-energy sector and then onto the fixed-particle-number subspace appropriate to the atomic system under consideration. 
More precisely, one projects the field algebra onto the corresponding sector of a fermionic Fock space with the prescribed numbers of electrons and nuclei, so that the effective degrees of freedom are represented by Pauli-type wave functions, together with the associated projected operators $X \rightarrow \hat{X}$ and the usual Newton–Wigner (and Born) notion of localization~\cite{Newton,Wightman,Schroer}, under which the wave function admits the standard probabilistic interpretation. 
In general, such a reduction is observer dependent, and different observers can disagree on the resulting non-relativistic particle content~\cite{Unruh,Letaw2}.
This frame dependence is controlled by the observer's proper acceleration~\cite{Falcone3} and becomes negligible in the quasi-inertial regime of sufficiently small accelerations~\cite{Falcone3,Falcone4,Wald}. 
For a general atom, this projection is onto the sector with the appropriate electronic and nuclear content; the specific case of hydrogen is treated in Sec.~\ref{Sec:JaynesCummingsHydrogen}, where the projection is onto the one-electron/one-proton sector.

The quasi-inertial argument above was developed for linearly accelerated observers~\cite{Falcone3,Falcone4} and its extension to a rotating observer requires separate justification which we leave for future work; however, we believe the argument to hold for uniform rotation for the reasons below. 
Since we take $\mathbf{\Omega}$ to be constant, the generalized Born metric remains stationary $\rpd{t} g_{\mu \nu} = 0$ and retains the timelike Killing vector from Eq.~\eqref{KillingMag}.
It is this time-translation symmetry that permits an unambiguous, time-independent positive-energy projection for the Dirac field, just as it underlies the stationary field-mode decomposition used in Appendix~\ref{Appendix:EMquantization}.
Furthermore, uniform rotation is known to leave the natural creation and annihilation operators of a scalar field unchanged from their inertial-frame form, whereas non-uniform rotation induces genuine mode mixing and a new, populated vacuum state~\cite{Denardo,Letaw}; an analogous statement is expected to hold for Dirac fields~\cite{Obukhov}. 
Taking the observer to be uniformly and non-relativistically rotating with sufficiently small proper (centripetal) acceleration, their local frame is therefore quasi-inertial in the required sense, so rotating and inertial observers agree on the notion of an ``atom'' and on the particle content inside the cavity, with no new vacuum or associated Unruh-like excitations retained at the order considered.
This effective picture is also consistent with, though not established by, the localized-probe intuition underlying Unruh-DeWitt-type detector models~\cite{Wald,Lopp,Stritzelberger}.
With this, the leading-order Hamiltonian in the rotating frame, Eq.~\eqref{H_0}, is promoted to its operator form
\begin{equation} \label{H_0_proj}
    \hat{H}_0 = \frac{\left( \hat{\mathbf{p}} - q \hat{\mathbf{A}} \right)^2}{2 m} - \frac{g q}{2 m} \hat{\mathbf{S}} \bm{\cdot} \hat{\mathbf{B}} + q \hat{\varphi} - \mathbf{\Omega} \bm{\cdot} \left[ \hat{\mathbf{J}} - q \left( \hat{\mathbf{r}} \cross \hat{\mathbf{A}} \right) \right].
\end{equation}

\section{Modified Jaynes-Cummings Model for Hydrogen} \label{Sec:JaynesCummingsHydrogen}

\subsection{Center-of-mass coordinates} \label{Sec:CoMcoords}
Having worked out the non-relativistic limit of the Dirac equation and the E\&M field Hamiltonian in the generalized Born metric, we can now take Eqs.~\eqref{H_0_proj} and~\eqref{H_F} to build the interaction of hydrogen atoms with the E\&M field to lowest-order.
Specifically, we now include an electron/positron field $\Psi$ with mass $m_e$ and charge $q_e = - e$ and a proton/anti-proton field $\Phi$ with mass $m_p$ and charge $q_p = e$, such that the non-relativistic limit for protium (zero-neutron hydrogen) leads to the lowest-order atomic and atom-field interaction Hamiltonian
\begin{equation}
    \hat{H}_{\mathrm{A}} + \hat{H}_{\mathrm{AF}} = \hat{H}_0^e + \hat{H}_0^p,
\end{equation}
where $\hat{H}_0^i$ is the Hamiltonian from Eq.~\eqref{H_0} for the respective particle $i \in \{ e, p \}$.
The total Hamiltonian, now in SI units, is then given by
\begin{equation}
    \begin{aligned}
& \hat{H}_{\mathrm{H}} = \hat{H}_{\mathrm{F}} + \hat{H}_{\mathrm{A}} + \hat{H}_{\mathrm{AF}} \\
&= \int d^3 r \left[ \frac{\epsilon_0}{2} \left( \hat{E}^2 + c^2 \hat{B}^2 \right) - \mathbf{\Omega} \bm{\cdot} \hat{\bm{\ell}} \right] \\
& + \frac{\left( \hat{\mathbf{p}}_e + e \hat{\mathbf{A}} \right)^2}{2 m_e} + \frac{g_e e}{2 m_e} \hat{\mathbf{S}} \bm{\cdot} \hat{\mathbf{B}} - e \hat{\varphi} - \mathbf{\Omega} \bm{\cdot} \left[ \hat{\mathbf{J}}_e + e \left( \hat{\mathbf{r}}_e \cross \hat{\mathbf{A}} \right) \right] \\
& + \frac{\left( \hat{\mathbf{p}}_p - e \hat{\mathbf{A}} \right)^2}{2 m_p} - \frac{g_p e}{2 m_p} \hat{\mathbf{I}} \bm{\cdot} \hat{\mathbf{B}} + e \hat{\varphi} - \mathbf{\Omega} \bm{\cdot} \left[ \hat{\mathbf{J}}_p - e \left( \hat{\mathbf{r}}_p \cross \hat{\mathbf{A}} \right) \right],
    \end{aligned}
\end{equation}
where $\hat{\varphi}$, $\hat{\mathbf{A}}$, and $\hat{\mathbf{B}}$ in the third and fourth lines are evaluated at the electron's ($\hat{\mathbf{r}}_e$) and proton's ($\hat{\mathbf{r}}_p$) position, respectively. 
Here, we have specified the electron spin $\hat{\mathbf{S}} = \hbar \hat{\bm{\sigma}}_e / 2$ and the nuclear spin $\hat{\mathbf{I}} = \hbar \hat{\bm{\sigma}}_p / 2$, and one can now insert the experimental values of the spin $g$-factor for the electron $g_e$ and proton $g_p$ ($g_p$ significantly deviates from $2$ because of its quark substructure). 
Note that the integral in the field Hamiltonian here was written specifically in the generalized Born metric [Eq.~\eqref{H_F}] with $\sqrt{\gamma} = 1$; in general, a factor of $\sqrt{\gamma}$ appears in the spatial volume element $d \mathcal{V} = \sqrt{\gamma} d^3 r$ (playing the role of the Jacobian factor over the Cauchy hypersurface). 

As discussed in Ref.~\cite{Steck}, one traditionally proceeds in the Minkowski metric by decomposing the electric field using Helmholtz theorem~\cite{Griffiths_EM,Steck} $\hat{\mathbf{E}}' = \hat{\mathbf{E}}'_{\parallel} + \hat{\mathbf{E}}'_{\perp}$ with $\bm{\nabla}' \cross \hat{\mathbf{E}}'_{\parallel} = 0$ and $\bm{\nabla}' \bm{\cdot} \hat{\mathbf{E}}'_{\perp} = 0$.
One then associates the longitudinal (curl-free) field $\hat{\mathbf{E}}'_{\parallel} = - \bm{\nabla}' \hat{\varphi}'$ and the $\pm e \hat{\varphi}'$ terms in the Hamiltonian with the binding potential $V(\hat{\mathbf{r}}')$ from the Coulomb interaction within the atom $\hat{\mathbf{E}}'_{\mathrm{I}} = \hat{\mathbf{E}}'_{\parallel}$, while the transverse (divergence-free) field $\hat{\mathbf{E}}'_{\perp} = - \rpd{T} \hat{\mathbf{A}}'$ represents the free field $\hat{\mathbf{E}}'_{\mathrm{F}} = \hat{\mathbf{E}}'_{\perp}$.
Importantly, in the Coulomb gauge, one then has $\hat{\mathbf{E}}'_{\mathrm{I}} \bm{\cdot} \hat{\mathbf{E}}'_{\mathrm{F}} = 0$ upon integrating by parts and dropping boundary terms, and so $\hat{E}'^2 = \hat{E}'^2_{\mathrm{I}} + \hat{E}'^2_{\mathrm{F}} = V(\hat{\mathbf{r}}') + \hat{E}_{\perp}^2$. 
In the generalized Born metric, however, one must be more careful.
Defining the internal and free fields as
\begin{equation}
    \hat{\mathbf{E}} = \hat{\mathbf{E}}_{\mathrm{I}} + \hat{\mathbf{E}}_{\mathrm{F}} = \left[ - \bm{\nabla} \hat{\varphi} \right] + \left[ - \rpd{t} \hat{\mathbf{A}} - \left( \mathbf{\Omega} \cross \mathbf{r} \right) \cross \hat{\mathbf{B}} \right],
\end{equation}
we see that the free electric field now has a longitudinal component using vector calculus identities~\cite{Jackson} (and $\bm{\nabla} \bm{\cdot} \mathbf{\Omega} = \bm{\nabla} \mathbf{\Omega} = 0$),
\begin{equation}
    \begin{aligned}
\bm{\nabla} \bm{\cdot} \hat{\mathbf{E}}_{\mathrm{F}} &= - \bm{\nabla} \bm{\cdot} \left[ \left( \mathbf{\Omega} \cross \mathbf{r} \right) \cross \hat{\mathbf{B}} \right] \\
&= \left( \bm{\nabla} \cross \hat{\mathbf{B}} \right) \bm{\cdot} \left( \mathbf{\Omega} \cross \mathbf{r} \right) - 2 \mathbf{\Omega} \bm{\cdot} \hat{\mathbf{B}},
    \end{aligned}
\end{equation}
due to the modification of Maxwell's equations from Sec.~\ref{Sec:FieldHamiltonian}~\footnote{The free electric field becomes transverse when $\mathbf{\Omega}$ is perpendicular to the plane of the cavity, such as the simplifications in Secs.~\ref{Sec:EMsimplify} and~\ref{Sec:SepVariables}}.
Thus, we expand
\begin{equation} \label{FullE2}
    \begin{aligned}
\hat{E}^2 &= \hat{E}_{\mathrm{I}}^2 + \hat{E}_{\mathrm{F}}^2 + \hat{\mathbf{E}}_{\mathrm{I}} \bm{\cdot} \hat{\mathbf{E}}_{\mathrm{F}} + \hat{\mathbf{E}}_{\mathrm{F}} \bm{\cdot} \hat{\mathbf{E}}_{\mathrm{I}} \\
&= \hat{E}_{\mathrm{I}}^2 + \hat{E}_{\mathrm{F}}^2 + 2 \left( \bm{\nabla} \hat{\varphi} \right) \bm{\cdot} \left[ \left( \mathbf{\Omega} \cross \mathbf{r} \right) \cross \hat{\mathbf{B}} \right],
    \end{aligned}
\end{equation}
where we have dropped divergence-free terms by integration by parts and used $[\bm{\nabla} \hat{\varphi}, \left( \mathbf{\Omega} \cross \mathbf{r} \right) \cross \hat{\mathbf{B}}] = 0$ since they are both solely functions of position. 
Meanwhile, the final term in the rotating field Hamiltonian Eq.~\eqref{H_F} leads to $\mathbf{\Omega} \bm{\cdot} \hat{\bm{\ell}} = \mathbf{\Omega} \bm{\cdot} \left( \hat{\bm{\ell}}_{\mathrm{I}} + \hat{\bm{\ell}}_{\mathrm{F}} \right)$ with $\hat{\bm{\ell}}_i = \epsilon_0 \mathbf{r} \cross (\hat{\mathbf{E}}_i \cross \hat{\mathbf{B}})$ (where $i \in \{ \mathrm{I}, \mathrm{F} \}$). 
Using the scalar triple product~\cite{Jackson}, we find
\begin{equation}
    \begin{aligned}
- \mathbf{\Omega} \bm{\cdot} \hat{\bm{\ell}}_{\mathrm{I}} &= \epsilon_0 \mathbf{\Omega} \bm{\cdot} \left( \mathbf{r} \cross \left[ \left( \bm{\nabla} \hat{\varphi} \right) \cross \hat{\mathbf{B}} \right] \right) \\
&= - \epsilon_0 \left( \bm{\nabla} \hat{\varphi} \right) \bm{\cdot} \left[ \left( \mathbf{\Omega} \cross \mathbf{r} \right) \cross \hat{\mathbf{B}} \right],
    \end{aligned}
\end{equation}
which perfectly cancels the contribution of the final term in Eq.~\eqref{FullE2} to the Hamiltonian (which is multiplied by an $\epsilon_0 / 2$).
Therefore, we can write
\begin{equation}
    \begin{aligned}
& \hat{H}_{\mathrm{H}} = \int d^3 r \left[ \frac{\epsilon_0}{2} \left( \hat{E}_{\mathrm{F}}^2 + c^2 \hat{B}^2 \right) - \mathbf{\Omega} \bm{\cdot} \hat{\bm{\ell}}_{\mathrm{F}} \right] + V(\hat{\mathbf{r}}_e - \hat{\mathbf{r}}_p) \\
& + \frac{\left( \hat{\mathbf{p}}_e + e \hat{\mathbf{A}} \right)^2}{2 m_e} + \frac{g_e e}{2 m_e} \hat{\mathbf{S}} \bm{\cdot} \hat{\mathbf{B}} - \mathbf{\Omega} \bm{\cdot} \left[ \hat{\mathbf{J}}_e + e \left( \hat{\mathbf{r}}_e \cross \hat{\mathbf{A}} \right) \right] \\
& + \frac{\left( \hat{\mathbf{p}}_p - e \hat{\mathbf{A}} \right)^2}{2 m_p} - \frac{g_p e}{2 m_p} \hat{\mathbf{I}} \bm{\cdot} \hat{\mathbf{B}} - \mathbf{\Omega} \bm{\cdot} \left[ \hat{\mathbf{J}}_p - e \left( \hat{\mathbf{r}}_p \cross \hat{\mathbf{A}} \right) \right],
    \end{aligned}
\end{equation}
where $\hat{\mathbf{A}}$ and $\hat{\mathbf{B}}$ in the second and third lines are evaluated at the electron's ($\hat{\mathbf{r}}_e$) and proton's ($\hat{\mathbf{r}}_p$) position, respectively.
Here, we have noted that the internal Coulomb interaction $V(\hat{\mathbf{r}})$ only depends on the relative distance between the two particles.

Next, we change to center-of-mass (CoM) coordinates of the atom using~\cite{Goldstein,Steck}
\begin{equation} \label{CoM_xp}
    \begin{aligned}
\hat{\mathbf{R}} = \frac{m_e \hat{\mathbf{r}}_e + m_p \hat{\mathbf{r}}_p}{M}, \quad \hat{\mathbf{P}} = \hat{\mathbf{p}}_e + \hat{\mathbf{p}}_p, \\
\tilde{\hat{\mathbf{r}}}_{\alpha} = \hat{\mathbf{r}}_{\alpha} - \hat{\mathbf{R}}, \quad \tilde{\hat{\mathbf{p}}}_{\alpha} = \hat{\mathbf{p}}_{\alpha} - \frac{m_{\alpha}}{M} \hat{\mathbf{P}},
    \end{aligned}
\end{equation}
with $\alpha \in \{ e, p \}$ and the total mass $M = m_e + m_p$. 
The top line represents the CoM coordinates while the second line represents the relative coordinates about the CoM. 
We also assume the total charge $Q = \sum_{\alpha} q_{\alpha}$ is zero throughout this article as we are interested in neutral atoms. 
Note that the CoM coordinates commute with the relative coordinates, and the CoM coordinates are canonical variables [see Eq.~\eqref{CommutationRelations}]. 
With this, we can expand the angular momentum of the two particles as
\begin{equation}
    \begin{aligned}
\hat{\mathbf{J}}_e + \hat{\mathbf{J}}_p &= \hat{\mathbf{r}}_e \cross \hat{\mathbf{p}}_e + \hat{\mathbf{S}} + \hat{\mathbf{r}}_p \cross \hat{\mathbf{p}}_p + \hat{\mathbf{I}} \\
&= \hat{\mathbf{F}} + \tilde{\hat{\mathbf{r}}}_p \cross \tilde{\hat{\mathbf{p}}}_p + \hat{\bm{\mathfrak{L}}},
    \end{aligned}
\end{equation}
where we have used $\sum_{\alpha} m_{\alpha} \tilde{\hat{\mathbf{r}}}_{\alpha} = 0$ and $\sum_{\alpha} \tilde{\hat{\mathbf{p}}}_{\alpha} = 0$, identified the orbital angular momentum of the electron $\hat{\mathbf{L}} = \tilde{\hat{\mathbf{r}}}_e \cross \tilde{\hat{\mathbf{p}}}_e$ and CoM angular momentum $\hat{\bm{\mathfrak{L}}} = \hat{\mathbf{R}} \cross \hat{\mathbf{P}}$, and defined the usual total angular momenta $\hat{\mathbf{J}} = \hat{\mathbf{L}} + \hat{\mathbf{S}}$ and $\hat{\mathbf{F}} = \hat{\mathbf{J}} + \hat{\mathbf{I}}$. 
Since the nucleus is much heavier than the electron $m_p \gg m_e$, we assume the CoM is near the proton's (quantum) position such that $\hat{\mathbf{R}} \approx \hat{\mathbf{r}}_p$ and we can ignore the angular momentum of the nucleus around the CoM,
\begin{equation}
    \hat{\mathbf{J}}_e + \hat{\mathbf{J}}_p \approx \hat{\mathbf{F}} + \hat{\bm{\mathfrak{L}}}.
\end{equation}
Plugging the CoM coordinates into the other terms in the Hamiltonian, we obtain
\begin{equation} \label{H_CoM}
    \begin{aligned}
& \hat{H}_{\mathrm{H}} = \int d^3 r \left[ \frac{\epsilon_0}{2} \left( \hat{E}_{\mathrm{F}}^2 + c^2 \hat{B}^2 \right) - \mathbf{\Omega} \bm{\cdot} \hat{\bm{\ell}}_{\mathrm{F}} \right] \\
& + V(\tilde{\hat{\mathbf{r}}}_e - \tilde{\hat{\mathbf{r}}}_p) + \frac{g_e e}{2 m_e} \hat{\mathbf{S}} \bm{\cdot} \hat{\mathbf{B}} (\hat{\mathbf{r}}_e) - \frac{g_p e}{2 m_p} \hat{\mathbf{I}} \bm{\cdot} \hat{\mathbf{B}} (\hat{\mathbf{r}}_p) \\
& + \frac{\hat{\mathbf{P}}^2}{2 M} + \frac{\left[ \tilde{\hat{\mathbf{p}}}_e + e \hat{\mathbf{A}} (\hat{\mathbf{r}}_e) \right]^2}{2 m_e} + \frac{\left[ \tilde{\hat{\mathbf{p}}}_p - e \hat{\mathbf{A}} (\hat{\mathbf{r}}_p) \right]^2}{2 m_p} \\
& + \frac{e}{2 M} \left( \hat{\mathbf{P}} \bm{\cdot} \left[ \hat{\mathbf{A}} (\hat{\mathbf{r}}_e) - \hat{\mathbf{A}} (\hat{\mathbf{r}}_p) \right] + \left[ \hat{\mathbf{A}} (\hat{\mathbf{r}}_e) - \hat{\mathbf{A}} (\hat{\mathbf{r}}_p) \right] \bm{\cdot} \hat{\mathbf{P}} \right) \\
& - \mathbf{\Omega} \bm{\cdot} \left( \hat{\mathbf{F}} + \hat{\bm{\mathfrak{L}}} \right) + e \mathbf{\Omega} \bm{\cdot} \left[ \tilde{\hat{\mathbf{r}}}_p \cross \hat{\mathbf{A}} (\hat{\mathbf{r}}_p) - \tilde{\hat{\mathbf{r}}}_e \cross \hat{\mathbf{A}} (\hat{\mathbf{r}}_e) \right] \\
& + e \mathbf{\Omega} \bm{\cdot} \left( \hat{\mathbf{R}} \cross \left[ \hat{\mathbf{A}} \left( \hat{\mathbf{r}}_p \right) - \hat{\mathbf{A}} \left( \hat{\mathbf{r}}_e \right) \right] \right).
    \end{aligned}
\end{equation}

\subsection{Power-Zienau-Woolley transformation}
Next, we can transform from a minimal coupling representation to a multipole expansion representation by performing a Power-Zienau-Woolley (PZW) transformation~\cite{Power,Power2,Power3,Power4,Woolley,Woolley2,Babiker,Babiker2,CohenTannoudji,Andrews,Steck}. 
Here, we take the atomic charge density as~\cite{Woolley2}
\begin{equation}
    \rho (\mathbf{r}) = e \delta^{(3)} \left( \mathbf{r} - \hat{\mathbf{r}}_p \right) - e \delta^{(3)} \left( \mathbf{r} - \hat{\mathbf{r}}_e \right),
\end{equation}
which, choosing a (non-unique) Wilson path~\cite{Wilson} along $\tilde{\hat{\mathbf{r}}}_{\alpha}$, gives the atomic polarization~\cite{Woolley2,Steck}
\begin{equation}
    \hat{\bm{\mathcal{P}}} (\mathbf{r}) = \sum_{\alpha} q_{\alpha} \tilde{\hat{\mathbf{r}}}_{\alpha} \int_0^1 ds \delta^{(3)} \left( \mathbf{r} - \hat{\mathbf{R}} - s \tilde{\hat{\mathbf{r}}}_{\alpha} \right),
\end{equation}
through its relation to the bound charge density~\cite{Griffiths_EM}
\begin{equation}
    \hat{\bm{\nabla}} \bm{\cdot} \hat{\bm{\mathcal{P}}} = - \rho. 
\end{equation}
From this, one can obtain an electric multipole series by expanding the Dirac delta functions about $s = 0$ (technically, expanding a test function that the delta function acts on~\cite{Steck}),
\begin{equation}
    \begin{aligned}
\delta^{(3)} \left( \mathbf{r} - \hat{\mathbf{R}} - s \tilde{\hat{\mathbf{r}}}_{\alpha} \right) = & \delta^{(3)}(\mathbf{r} - \hat{\mathbf{R}}) - s (\tilde{\hat{\mathbf{r}}}_{\alpha} \bm{\cdot} \hat{\bm{\nabla}}) \delta^{(3)} (\mathbf{r} - \hat{\mathbf{R}}) \\
& + \frac{s^2}{2} (\tilde{\hat{\mathbf{r}}}_{\alpha} \bm{\cdot} \hat{\bm{\nabla}})^2 \delta^{(3)} (\mathbf{r} - \hat{\mathbf{R}}) + \ldots,
    \end{aligned} 
\end{equation}
such that the polarization vector becomes~\cite{Steck}
\begin{equation} \label{Pexpansion}
    \begin{aligned}
\hat{\bm{\mathcal{P}}} (\mathbf{r}) = & \sum_{\alpha} q_{\alpha} \tilde{\hat{\mathbf{r}}}_{\alpha} \left[ \delta^{(3)} (\mathbf{r} - \hat{\mathbf{R}}) - \frac{1}{2} (\tilde{\hat{\mathbf{r}}}_{\alpha} \bm{\cdot} \hat{\bm{\nabla}}) \delta^{(3)} (\mathbf{r} - \hat{\mathbf{R}}) \right. \\
& \left. + \frac{1}{6} (\tilde{\hat{\mathbf{r}}}_{\alpha} \bm{\cdot} \hat{\bm{\nabla}})^2 \delta^{(3)} (\mathbf{r} - \hat{\mathbf{R}}) + \ldots \right],
    \end{aligned}
\end{equation}
whose terms will lead to an electric dipole moment, quadrupole moment, octupole moment, and so on. 
With this, we can define the PZW transformation through the unitary operator
\begin{equation} \label{U_PZW}
    \hat{U}_{\mathrm{PZW}} = \exp \left[ - \frac{i}{\hbar} \int d^3 r \hat{\bm{\mathcal{P}}} (\mathbf{r}) \bm{\cdot} \hat{\mathbf{A}} (\mathbf{r}) \right].
\end{equation}
Note that when the atomic polarization is defined with respect to a single reference point as is done here, the PZW transformation can be considered as a gauge change from the Coulomb gauge to the Poincar\'e gauge~\cite{CohenTannoudji}.

We are interested in optical wavelengths $\mathcal{O}(100 \, \mathrm{nm})$ while the size of atoms are typically $\mathcal{O}(\mathring{A}) = \mathcal{O}(0.1 \, \mathrm{nm})$, and so we typically can invoke the long-wavelength approximation in which we assume the E\&M field is constant over the size of the atom, which amounts to setting $\hat{\mathbf{A}} (\hat{\mathbf{r}}_{\alpha}) \approx \hat{\mathbf{A}} (\hat{\mathbf{R}})$ and similarly for $\hat{\mathbf{E}}$ and $\hat{\mathbf{B}}$. 
Moreover, we make the dipole approximation which amounts to picking out the first term in the polarization's expansion Eq.~\eqref{Pexpansion},
\begin{equation}
    \hat{\bm{\mathcal{P}}} (\mathbf{r}) \approx \sum_{\alpha} q_{\alpha} \tilde{\hat{\mathbf{r}}}_{\alpha} \delta^{(3)} (\mathbf{r} - \hat{\mathbf{R}}) = \hat{\mathbf{d}} \delta^{(3)} (\mathbf{r} - \hat{\mathbf{R}}), 
\end{equation}
with the electric dipole operator
\begin{equation}
    \hat{\mathbf{d}} = \sum_{\alpha} q_{\alpha} \tilde{\hat{\mathbf{r}}}_{\alpha}. 
\end{equation}
The full transformation without making the dipole and long-wavelength approximations is considered in, e.g., Refs.~\cite{Steck,Power4,Baxter,Lembessis,Andrews,Vukics,Kattan}.

We can now work out how the various terms in the Hamiltonian transform under $\hat{U}_{\mathrm{PZW}}$. 
The transformations are typically found by utilizing the Baker-Campbell-Hausdorff formula~\cite{Bonfiglioli,Sakurai}, 
\begin{equation}
    e^{i \hat{G} \upsilon} \hat{O} e^{- i \hat{G} \upsilon} = \hat{O} + i \upsilon \left[ \hat{G}, \hat{O} \right] + \frac{(i \upsilon)^2}{2!} \left[ \hat{G}, \left[ \hat{G}, \hat{O} \right] \right] + \ldots,
\end{equation}
with real parameter $\upsilon$. 
Thus, we require the commutation relations between the operators of interest and the $\hat{\bm{\mathcal{P}}}$ and $\hat{\mathbf{A}}$ in $\hat{U}_{\mathrm{PZW}}$.
Of the operators in Eq.~\eqref{H_CoM}, the fundamental non-zero commutators are given by~\cite{Steck,Power3,Baxter,Lembessis,Barnett,Gardiner,Blow} 
\begin{equation} \label{CommutationRelations}
    \begin{aligned}
\left[ \tilde{\hat{\mathbf{p}}}_{\alpha,j}, \tilde{\hat{\mathbf{r}}}_{\beta,k} \right] &= - i \hbar \delta_{jk} \left( \delta_{\alpha \beta} - \frac{m_{\alpha}}{M} \right), \\
\left[ \hat{\mathbf{P}}_j, \hat{\mathbf{R}}_k \right] &= - i \hbar \delta_{j k}, \\
\left[ \hat{\mathbf{A}}_j (\mathbf{r},t), \hat{\mathbf{E}}_{\mathrm{F},k} (\mathbf{r}',t) \right] &= - \frac{i \hbar}{\epsilon_0} \delta_{j k}^{\perp} \left( \mathbf{r} - \mathbf{r}' \right),
    \end{aligned}
\end{equation}
where the indices here indicate vector components rather than explicit co- or contravariant indices, while the transverse delta function $\delta^{\perp} (\mathbf{r} - \mathbf{r}')$ serves as the projection kernel onto the space of divergence-free vector fields in the Coulomb gauge [see Eq.~\eqref{A_Pi_comm}].
Note that the commutator of the vector potential and electric field is the same as in Minkowksi spacetime because $[\hat{\mathbf{A}}_j, [(\mathbf{\Omega} \cross \mathbf{r}) \cross \hat{\mathbf{B}}]_k ] = 0$.
Also note that in the field commutator, we have ignored contributions of induced longitudinal modes derived in Ref.~\cite{Power3} as this term, after the PWZ transformation, exactly cancels~\cite{Power3} a contribution from an image charge \emph{outside} the cavity due to the atomic Coulomb interaction inside the cavity which we had ignored in Eq.~\eqref{H_CoM}. 
We thus find that all second-order commutators for the operators of interest $\hat{O} \in \{ \tilde{\hat{\mathbf{p}}}_{\alpha}, \hat{\mathbf{P}}, \hat{\mathbf{E}}_{\mathrm{F}} \}$ are zero such that
\begin{equation}
    \hat{U}_{\mathrm{PZW}} \hat{O} \hat{U}_{\mathrm{PZW}}^{\dagger} = \hat{O} - \frac{i}{\hbar} \int d^3 r \left[ \hat{\bm{\mathcal{P}}} (\mathbf{r}) \bm{\cdot} \hat{\mathbf{A}} (\mathbf{r}), \hat{O} \right].
\end{equation}
We then obtain the PWZ transformation on the CoM and relative momenta under the dipole and long-wavelength approximations~\cite{Steck,Baxter,Lembessis},
\begin{equation} \label{MomTranform}
    \begin{aligned}
\hat{U}_{\mathrm{PZW}} \tilde{\hat{\mathbf{p}}}_{\alpha} \hat{U}_{\mathrm{PZW}}^{\dagger} &\approx \tilde{\hat{\mathbf{p}}}_{\alpha} + q_{\alpha} \hat{\mathbf{A}} (\hat{\mathbf{R}}) + \frac{q_{\alpha}}{2} \tilde{\hat{\mathbf{r}}}_{\alpha} \cross \hat{\mathbf{B}} (\hat{\mathbf{R}}), \\
\hat{U}_{\mathrm{PZW}} \hat{\mathbf{P}} \hat{U}_{\mathrm{PZW}}^{\dagger} &\approx \hat{\mathbf{P}} + \hat{\mathbf{d}} \cross \hat{\mathbf{B}} (\hat{\mathbf{R}}),
    \end{aligned}
\end{equation}
such that
\begin{widetext}
\begin{equation} \label{KEtransform}
    \begin{aligned}
& \hat{U}_{\mathrm{PZW}} \left( \sum_{\alpha} \frac{1}{2 m_{\alpha}} \left[ \tilde{\hat{\mathbf{p}}}_{\alpha} - q_{\alpha} \hat{\mathbf{A}} \right]^2 \right) \hat{U}_{\mathrm{PZW}}^{\dagger} \approx \frac{\tilde{\hat{\mathbf{p}}}_p^2}{2 m_p} + \frac{\tilde{\hat{\mathbf{p}}}_e^2}{2 m_e} - \hat{\mathbf{m}} \bm{\cdot} \hat{\mathbf{B}} + \frac{1}{8 m_e} \left[ \tilde{\hat{\mathbf{r}}}_e \cross \hat{\mathbf{B}} \right]^2 + \frac{1}{8 m_p} \left[ \tilde{\hat{\mathbf{r}}}_p \cross \hat{\mathbf{B}} \right]^2, \\
& \hat{U}_{\mathrm{PZW}} \frac{\hat{\mathbf{P}}^2}{2 M} \hat{U}_{\mathrm{PZW}}^{\dagger} \approx \frac{\hat{\mathbf{P}}^2}{2 M} + \frac{1}{2 M} \left[ \hat{\mathbf{d}} \cross \hat{\mathbf{B}} \right]^2 + \frac{1}{2 M} \left( \hat{\mathbf{P}} \bm{\cdot} \left[ \hat{\mathbf{d}} \cross \hat{\mathbf{B}} \right] + \left[ \hat{\mathbf{d}} \cross \hat{\mathbf{B}} \right] \bm{\cdot} \hat{\mathbf{P}} \right),
    \end{aligned}
\end{equation}
\end{widetext}
where all fields are evaluated at $\hat{\mathbf{R}}$ and we have defined the magnetic dipole operator
\begin{equation}
    \hat{\mathbf{m}} = \sum_{\alpha} \frac{q_{\alpha}}{2 m_{\alpha}} \left( \tilde{\hat{\mathbf{r}}}_{\alpha} \cross \tilde{\hat{\mathbf{p}}}_{\alpha} \right). 
\end{equation}
Meanwhile, the commutator of the electric field leads to~\cite{Steck}
\begin{equation} \label{Etransform}
    \hat{U}_{\mathrm{PZW}} \hat{\mathbf{E}}_{\mathrm{F}} (\mathbf{r}) \hat{U}_{\mathrm{PZW}}^{\dagger} = \hat{\mathbf{E}}_{\mathrm{F}} (\mathbf{r}) - \frac{1}{\epsilon_0} \hat{\bm{\mathcal{P}}}_{\perp} (\mathbf{r}),
\end{equation}
which has the form of the displacement field operator~\cite{Griffiths_EM}, where $\hat{\bm{\mathcal{P}}}_{\perp}$ is used to distinguish the transverse polarization from the longitudinal polarization $\hat{\bm{\mathcal{P}}}_{\parallel}$ which is already included in the Coulomb binding potential~\cite{CohenTannoudji,Steck}. 
This then gives
\begin{equation} \label{E2transform}
    \begin{aligned}
\hat{U}_{\mathrm{PZW}} \left[ \hat{\mathbf{E}}_{\mathrm{F}} (\mathbf{r}) \right]^2 \hat{U}_{\mathrm{PZW}}^{\dagger} = & \left[ \hat{\mathbf{E}}_{\mathrm{F}} (\mathbf{r}) \right]^2 + \frac{1}{\epsilon_0^2} \left[ \hat{\bm{\mathcal{P}}}_{\perp} (\mathbf{r}) \right]^2 \\
& - \frac{2}{\epsilon_0} \hat{\bm{\mathcal{P}}}_{\perp} (\mathbf{r}) \bm{\cdot} \hat{\mathbf{E}}_{\mathrm{F}} (\mathbf{r}) \\
\approx & \left[ \hat{\mathbf{E}}_{\mathrm{F}} (\mathbf{r}) \right]^2 + \frac{1}{\epsilon_0^2} \left[ \hat{\bm{\mathcal{P}}}_{\perp} (\mathbf{r}) \right]^2 \\
& - \frac{2}{\epsilon_0} \hat{\mathbf{d}} \bm{\cdot} \hat{\mathbf{E}}_{\mathrm{F}} (\mathbf{r}) \delta^{(3)} (\mathbf{r} - \hat{\mathbf{R}}),
    \end{aligned}
\end{equation}
where we have used $[\hat{\bm{\mathcal{P}}}_{\perp}, \hat{\mathbf{E}}_{\mathrm{F}}] = 0$.
We also obtain
\begin{equation} \label{l_EM_PZtransform}
    \begin{aligned}
\hat{U}_{\mathrm{PZW}} \hat{\bm{\ell}}_{\mathrm{F}} (\mathbf{r}) \hat{U}_{\mathrm{PZW}}^{\dagger} &= \epsilon_0 \mathbf{r} \cross \left[ \left( \hat{U}_{\mathrm{PZW}} \hat{\mathbf{E}}_{\mathrm{F}} (\mathbf{r}) \hat{U}_{\mathrm{PZW}}^{\dagger} \right) \cross \hat{\mathbf{B}} (\mathbf{r}) \right] \\
&= \hat{\bm{\ell}}_{\mathrm{F}} (\mathbf{r}) - \mathbf{r} \cross \left[ \hat{\bm{\mathcal{P}}}_{\perp} (\mathbf{r}) \cross \hat{\mathbf{B}} (\mathbf{r}) \right] \\
&\approx \hat{\bm{\ell}}_{\mathrm{F}} (\mathbf{r}) - \mathbf{r} \cross \left[ \hat{\mathbf{d}} \cross \hat{\mathbf{B}} (\mathbf{r}) \right] \delta^{(3)}(\mathbf{r} - \hat{\mathbf{R}}).
    \end{aligned}
\end{equation}
Finally, we utilize Eq.~\eqref{MomTranform} to find
\begin{equation} \label{AngMomTransform}
    \begin{aligned}
& \hat{U}_{\mathrm{PZW}} \hat{\mathbf{F}} \hat{U}_{\mathrm{PZW}}^{\dagger} \approx \hat{\mathbf{F}} + \hat{\mathbf{d}} \cross \hat{\mathbf{A}} (\hat{\mathbf{R}}) + \frac{\tilde{\hat{\mathbf{r}}}_e}{2} \cross \left[ \hat{\mathbf{d}} \cross \hat{\mathbf{B}} (\hat{\mathbf{R}}) \right], \\
& \hat{U}_{\mathrm{PZW}} \hat{\bm{\mathfrak{L}}} \hat{U}_{\mathrm{PZW}}^{\dagger} \approx \hat{\bm{\mathfrak{L}}} + \hat{\mathbf{R}} \cross \left[ \hat{\mathbf{d}} \cross \hat{\mathbf{B}} (\hat{\mathbf{R}}) \right].
    \end{aligned}
\end{equation}
Then using the scalar triple product identity~\cite{Jackson}, we obtain
\begin{equation}
    \begin{aligned}
\mathbf{\Omega} \bm{\cdot} \left( \tilde{\hat{\mathbf{r}}}_e \cross \left[ \hat{\mathbf{d}} \cross \hat{\mathbf{B}} (\hat{\mathbf{R}}) \right] \right) &= \left[ \mathbf{\Omega} \cross \tilde{\hat{\mathbf{r}}}_e \right] \bm{\cdot} \left[ \hat{\mathbf{d}} \cross \hat{\mathbf{B}} (\hat{\mathbf{R}}) \right] \\
\mathbf{\Omega} \bm{\cdot} \left( \hat{\mathbf{R}} \cross \left[ \hat{\mathbf{d}} \cross \hat{\mathbf{B}} (\hat{\mathbf{R}}) \right] \right) &= \left[ \mathbf{\Omega} \cross \hat{\mathbf{R}} \right] \bm{\cdot} \left[ \hat{\mathbf{d}} \cross \hat{\mathbf{B}} (\hat{\mathbf{R}}) \right].
    \end{aligned}
\end{equation}

We now plug Eqs.~\eqref{MomTranform}--\eqref{KEtransform} and Eqs.~\eqref{Etransform}--\eqref{AngMomTransform} back into the Hamiltonian Eq.~\eqref{H_CoM} under the dipole and long-wavelength approximations, i.e., we set $\hat{\mathbf{A}} (\hat{\mathbf{r}}_{\alpha}) \approx \hat{\mathbf{A}} (\hat{\mathbf{R}})$ throughout Eq.~\eqref{H_CoM}.
Furthermore, after inserting these equations into Eq.~\eqref{H_CoM}, we drop any terms proportional to $\tilde{\hat{\mathbf{r}}}_p$ or $\tilde{\hat{\mathbf{p}}}_p$ as we again assume $m_e \ll m_p$ such that $\hat{\mathbf{r}}_p \approx \hat{\mathbf{R}}$ and $\hat{\mathbf{p}}_p \approx \hat{\mathbf{P}}$. 
With this approximation, note that $V(\tilde{\hat{\mathbf{r}}}_e - \tilde{\hat{\mathbf{r}}}_p) \approx V(\tilde{\hat{\mathbf{r}}}_e)$, $\hat{\mathbf{d}} \approx - e \tilde{\hat{\mathbf{r}}}_e$, and $\hat{\mathbf{m}} \approx - e (\tilde{\hat{\mathbf{r}}}_e \cross \tilde{\hat{\mathbf{p}}}_e) / (2 m_e)$. 
The full PZW transformed Hamiltonian for hydrogen is then given by
\begin{widetext}
\begin{equation} \label{H_H}
    \begin{aligned}
\tilde{\hat{H}}_{\mathrm{H}} \approx & \frac{\hat{\mathbf{P}}^2}{2 M} + \frac{\tilde{\hat{\mathbf{p}}}_e^2}{2 m_e} + V \left( \tilde{\hat{\mathbf{r}}}_e \right) + \frac{\epsilon_0}{2} \int d^3 r \left( \hat{E}_{\mathrm{F}}^2 + c^2 \hat{B}^2 \right) + \frac{e}{2} \left[ \frac{g_e}{m_e} \hat{\mathbf{S}} - \frac{g_p}{m_p} \hat{\mathbf{I}} \right] \bm{\cdot} \hat{\mathbf{B}} (\hat{\mathbf{R}}) \\
& - \hat{\mathbf{d}} \bm{\cdot} \hat{\mathbf{E}}_{\mathrm{F}} (\hat{\mathbf{R}}) + \frac{1}{2 \epsilon_0} \int d^3 r \left[ \hat{\bm{\mathcal{P}}}_{\perp} (\mathbf{r}) \right]^2 - \hat{\mathbf{m}} \bm{\cdot} \hat{\mathbf{B}} (\hat{\mathbf{R}}) + \frac{1}{8 m_e} \left[ \hat{\mathbf{d}} \cross \hat{\mathbf{B}} (\hat{\mathbf{R}}) \right]^2 + \frac{1}{2 M} \left[ \hat{\mathbf{d}} \cross \hat{\mathbf{B}} (\hat{\mathbf{R}}) \right]^2 \\
& + \frac{1}{2 M} \left( \hat{\mathbf{P}} \bm{\cdot} \left[ \hat{\mathbf{d}} \cross \hat{\mathbf{B}} (\hat{\mathbf{R}}) \right] + \left[ \hat{\mathbf{d}} \cross \hat{\mathbf{B}} (\hat{\mathbf{R}}) \right] \bm{\cdot} \hat{\mathbf{P}} \right) \\
& - \mathbf{\Omega} \bm{\cdot} \left( \hat{\mathbf{F}} + \hat{\bm{\mathfrak{L}}} \right) - \frac{1}{2} \left[ \mathbf{\Omega} \cross \tilde{\hat{\mathbf{r}}}_e \right] \bm{\cdot} \left[ \hat{\mathbf{d}} \cross \hat{\mathbf{B}} (\hat{\mathbf{R}}) \right] - \int d^3 r \mathbf{\Omega} \bm{\cdot} \hat{\bm{\ell}}_{\mathrm{F}},
    \end{aligned}
\end{equation}
\end{widetext}
where a $\mathbf{\Omega} \bm{\cdot} [\hat{\mathbf{d}} \cross \hat{\mathbf{A}}(\hat{\mathbf{R}})]$ from Eq.~\eqref{H_CoM} and a $[\mathbf{\Omega} \cross \hat{\mathbf{R}}] \bm{\cdot} [\hat{\mathbf{d}} \cross \hat{\mathbf{B}} (\hat{\mathbf{R}})]$ from Eq.~\eqref{l_EM_PZtransform} have canceled the equivalent terms from Eq.~\eqref{AngMomTransform} and we have set the last line of Eq.~\eqref{H_CoM} to zero in the long-wavelength approximation.
We note in passing that the CoM coordinates used here are not the same as the center-of-energy coordinates used in Ref.~\cite{Sonnleitner3}.
Therefore, apparent friction forces can arise with these coordinates for moving atoms due to the fact that the total mass $M$ does not directly correspond to the mass of the atom as atomic mass also includes a contribution from the binding energy of the atom~\cite{Sonnleitner2,Sonnleitner3}.
Therefore, when considering the time-derivative of the atom's momentum, one must also include the change in mass from the binding energy. 
One can instead make a separate canonical transformation to center-of-energy coordinates after performing the PZW transformation which would unambiguously resolve these anomalous forces for ultra-precise experiments~\cite{Sonnleitner3}, but we do not pursue this here.

We can now go through and discuss the various terms in Eq.~\eqref{H_H}.
In the first line, we have the typical atomic Hamiltonian with the atom's total kinetic energy and internal structure,
\begin{equation}
    \begin{aligned}
\hat{H}_{\mathrm{A}} &= \hat{H}_{\mathrm{KE}} + \hat{H}_{\mathrm{AI}}, \\
\hat{H}_{\mathrm{KE}} = \frac{\hat{\mathbf{P}}^2}{2 M}, & \quad \hat{H}_{\mathrm{AI}} = \frac{\tilde{\hat{\mathbf{p}}}_e^2}{2 m_e} + V \left( \tilde{\hat{\mathbf{r}}}_e \right),
    \end{aligned}
\end{equation}
and then the typical E\&M field Hamiltonian in the inertial reference frame,
\begin{equation}
    \hat{H}_{\mathrm{IF}} = \frac{\epsilon_0}{2} \int d^3 r \left( \hat{E}_{\mathrm{F}}^2 + c^2 \hat{B}^2 \right).
\end{equation}
Meanwhile, the last term on the first line is the intrinsic magnetic moment of the electron and proton interacting with the magnetic field,
\begin{equation} \label{H_MM}
    \hat{H}_{\mathrm{MM}} = \frac{e}{2} \left[ \frac{g_e}{m_e} \hat{\mathbf{S}} - \frac{g_p}{m_p} \hat{\mathbf{I}} \right] \bm{\cdot} \hat{\mathbf{B}} (\hat{\mathbf{R}}).
\end{equation}
The second line of Eq.~\eqref{H_H} represents the atom-field interaction in the dipole approximation, which we split into the electric dipole interaction and electric self-energy of the dipole,
\begin{equation}
    \hat{H}_{\mathrm{AE}} = - \hat{\mathbf{d}} \bm{\cdot} \hat{\mathbf{E}}_{\mathrm{F}} (\hat{\mathbf{R}}) + \frac{1}{2 \epsilon_0} \int d^3 r \left[ \hat{\bm{\mathcal{P}}}_{\perp} (\mathbf{r}) \right]^2,
\end{equation}
and the magnetic dipole interaction and the diamagnetic interaction of the electric dipole,
\begin{equation}
    \hat{H}_{\mathrm{AM}} = - \hat{\mathbf{m}} \bm{\cdot} \hat{\mathbf{B}} (\hat{\mathbf{R}}) + \left( \frac{1}{8 m_e} + \frac{1}{2 M} \right) \left[ \hat{\mathbf{d}} \cross \hat{\mathbf{B}} (\hat{\mathbf{R}}) \right]^2.
\end{equation}
Next, the third line represents the R\"ontgen interaction~\cite{Rontgen} between a moving electric dipole and a magnetic field,
\begin{equation}
    \hat{H}_{\mathrm{R}} = \frac{1}{2 M} \left( \hat{\mathbf{P}} \bm{\cdot} \left[ \hat{\mathbf{d}} \cross \hat{\mathbf{B}} (\hat{\mathbf{R}}) \right] + \left[ \hat{\mathbf{d}} \cross \hat{\mathbf{B}} (\hat{\mathbf{R}}) \right] \bm{\cdot} \hat{\mathbf{P}} \right),
\end{equation}
which can be seen as an effective $\hat{\mathbf{d}} \bm{\cdot} \hat{\mathbf{E}}_{\mathrm{eff}}$ interaction~\cite{Lembessis,Baxter,Sonnleitner,Sonnleitner4,Steck}.
The fourth line represents the inertial effects on the atom and field which are artifacts of the rotating frame~\cite{Hehl}.
First, there is a spin-rotation coupling term~\cite{Mashhoon}, $\mathbf{\Omega} \bm{\cdot} (\hat{\mathbf{S}} + \hat{\mathbf{I}})$, and a Sagnac-like term~\cite{Rabi,Werner}, $\mathbf{\Omega} \bm{\cdot} (\hat{\mathbf{L}} + \hat{\bm{\mathfrak{L}}})$, which are both effective rotation-angular momentum couplings,
\begin{equation}
    \hat{H}_{\mathrm{RA}} = - \mathbf{\Omega} \bm{\cdot} \left( \hat{\mathbf{F}} + \hat{\bm{\mathfrak{L}}} \right).
\end{equation}
Then, there is a R\"ontgen interaction associated with the motion of the electron due to the rotation of an atom that is stationary in the non-inertial frame,
\begin{equation}
    \hat{H}_{\mathrm{RR}} = - \frac{1}{2} \left[ \mathbf{\Omega} \cross \tilde{\hat{\mathbf{r}}}_e \right] \bm{\cdot} \left[ \hat{\mathbf{d}} \cross \hat{\mathbf{B}} (\hat{\mathbf{R}}) \right].
\end{equation}
Finally, there is the Sagnac effect on the E\&M field from the rotation,
\begin{equation}
    \hat{H}_{\mathrm{SF}} = - \int d^3 r \mathbf{\Omega} \bm{\cdot} \hat{\bm{\ell}}_{\mathrm{F}}. 
\end{equation}

The inertial effects for the rotating observer are thus represented by $\hat{H}_{\mathrm{RA}}$, $\hat{H}_{\mathrm{RR}}$, and $\hat{H}_{\mathrm{SF}}$.
Traditional optical~\cite{Chow,Faucheux} and matter wave~\cite{Chih,LeDesma} gyroscopes rely on the Sagnac effect on the light $(- \mathbf{\Omega} \bm{\cdot} \hat{\bm{\ell}}_{\mathrm{F}})$ and the Sagnac-like effect on the CoM motion of the atoms $(- \mathbf{\Omega} \bm{\cdot} \hat{\bm{\mathfrak{L}}})$, respectively.
Above, we have derived two new lowest-order effects relevant for atom-cavity interactions with hydrogen:
\begin{enumerate}
    \item a rotation-induced hyperfine shift, 
    \begin{equation} \label{HyperfineShift}
        \hat{H}_{\mathrm{RH}} = - \mathbf{\Omega} \bm{\cdot} \hat{\mathbf{F}};
    \end{equation}

    \item a R\"ontgen interaction from the rotational motion of the dipole (distinct from its orbital angular momentum),
    \begin{equation} \label{RotationRontgen}
        \hat{H}_{\mathrm{RR}} = \frac{1}{2 e} \left[ \mathbf{\Omega} \cross \hat{\mathbf{d}} \right] \bm{\cdot} \left[ \hat{\mathbf{d}} \cross \hat{\mathbf{B}} (\hat{\mathbf{R}}) \right].
    \end{equation}
\end{enumerate}
While the R\"ontgen interaction is often quite small [see Eq.~\eqref{RotRontgenScale}], the rotational-induced hyperfine shift can be an appreciable shift that can impact quantum metrology experiments.
For example, upon generalizing our calculation to alkaline-earth(-like) atoms, we expect the frequency of a superradiant laser on a clock transition to be slaved to the atomic transition~\cite{Meiser,Reilly} such that $\Omega$ can be inferred directly by the rotational-induced hyperfine shift for an $\mathcal{O}(\mathrm{mHz})$ rotation, or possibly even smaller, which can be directly measured by simple heterodyne measurements~\cite{Reilly2}. 

Of course, if we had included the motion of the nucleus about the CoM, we would also obtain a Sagnac-like term and R\"ontgen interaction associated with its motion in the rotating frame. 
Furthermore, if we had included higher-order terms from the FW transformation we would see additional inertial effects arise which then leads to additional atom-cavity couplings. 
For example, the PZW transformation of the leading-order relativistic correction from Eq.~\eqref{H_r} is considered in Ref.~\cite{Asano}.

\subsection{Electric dipole approximation}
To simplify the model further, it is typical to next make the \emph{electric} dipole approximation.
Here, one notes that in a typical cavity system, one has $\abs{\mathbf{B}} \sim \abs{\mathbf{E}} / c$ and $\abs{\mathbf{d}} \sim e \text{\AA}$, such that (for low energy states of hydrogen~\cite{Bethe})
\begin{equation}
    \abs{\mathbf{m}} \abs{\mathbf{B}} / (\abs{\mathbf{d}} \abs{\mathbf{E}}) \sim \abs{\tilde{\mathbf{p}}_e} / (c m_e) \sim \abs{\tilde{\mathbf{v}}_e} / c \sim \alpha_{\mathrm{F}} \ll 1,
\end{equation}
with the fine structure constant $\alpha_{\mathrm{F}} \approx 1/137$, while
\begin{equation}
    \begin{aligned}
(\abs{\mathbf{d}} \abs{\mathbf{B}})^2 / (m_e \abs{\mathbf{d}} \abs{\mathbf{E}}) &\sim \abs{\mathbf{d}} \abs{\mathbf{B}} / (m_e c) \sim (e/m_e) \text{\AA} \abs{\mathbf{E}} / c^2 \\
&\sim (10 \; \mathrm{C} \cdot \mathrm{m} / \mathrm{kg}) \abs{\mathbf{E}} / c^2 \ll 1, \\
(\abs{\mathbf{d}} \abs{\mathbf{B}})^2 / (M \abs{\mathbf{d}} \abs{\mathbf{E}}) &\ll 1,
    \end{aligned}
\end{equation}
for non-relativistic E\&M fields.
Here, we have defined the relative electron velocity $\tilde{\mathbf{v}}_e = \tilde{\mathbf{p}}_e / m_e$, inserted the electron charge-to-mass ratio $e/m_e \sim 10^{11} \, \mathrm{C} / \mathrm{kg}$, and used $1/m_e \gg 1/M$.
Therefore, for typical E1 transitions, the magnetic dipole and diamagnetic interactions are much weaker than the electric dipole interaction such that they can be ignored; the magnetic dipole interaction is usually comparable with the electric quadrupole interaction~\cite{Steck,Griffiths_EM} while the diagmagnetic interactions are usually an order of $e \lambda \abs{\mathbf{E}} / (m_e c^2)$ smaller than this~\cite{Steck}. 
Similarly, we can compare the typical R\"ontgen interaction,
\begin{equation}
    \abs{\mathbf{P}} \abs{\mathbf{d}} \abs{\mathbf{B}} / (M \abs{\mathbf{d}} \abs{\mathbf{E}}) \sim \abs{\mathbf{P}} / (c M) \sim \abs{\mathbf{v}_{\mathrm{CoM}}} / c \ll 1,
\end{equation}
where $\mathbf{v}_{\mathrm{CoM}} = \mathbf{P} / M$, and the rotation-induced R\"ontgen interaction,
\begin{equation} \label{RotRontgenScale}
    \abs{\mathbf{\Omega}} \abs{\tilde{\mathbf{r}}_e} \abs{\mathbf{d}} \abs{\mathbf{B}} / (\abs{\mathbf{d}} \abs{\mathbf{E}}) \sim \abs{\mathbf{\Omega}} \text{\AA} / c \ll 1,
\end{equation}
and so these interactions can also be ignored for typical E1 transitions. 
While we do drop the R\"ontgen interaction based on this argument, it is important to note that the R\"ontgen interaction is in fact subtly required both for energy-momentum conservation and for radiation-induced mechanical forces to be gauge invariant~\cite{Baxter,Lopp,Sonnleitner}.
Furthermore, there have been various works that study the impact, and necessity, of the R\"ontgen interaction for modeling spontaneous emission from CoM motion of atoms~\cite{Wilkens,Wilkens2,Boussiakou,Cresser}, even at non-relativistic velocities, as well as resolving apparent friction forces in quantum optics from changes in the internal energy of atoms~\cite{Sonnleitner2,Sonnleitner3}. 

Next, for a localized atoms, the electric self-energy of the dipole is typically absorbed into a renormalized Hamiltonian for the atomic internal structure rather than treated as a dynamical light-matter interaction~\cite{Steck,Rokaj,Schafer}.
Accordingly, we define the renormalized atomic internal Hamiltonian for the bare states of the atom,
\begin{equation} \label{H_AIp}
    \hat{H}'_{\mathrm{AI}} = \hat{H}_{\mathrm{AI}} + \frac{1}{2 \epsilon_0} \int d^3 r \left[ \hat{\bm{\mathcal{P}}}_{\perp} (\mathbf{r}) \right]^2 + \tilde{\hat{H}}_{\mathrm{H,r}},
\end{equation}
and so the dipole self-energy effects are already incorporated into the physical atomic spectrum in our model. 
Here, $\tilde{\hat{H}}_{\mathrm{H,r}}$ represents relativistic corrections considered in, e.g., Refs.~\cite{Asano,Wundt2,Pachucki} applied to the protium atom, such as the Darwin term, spin-orbit coupling, and mass defects.
We note that the modification of the free electric field in the rotating frame, $- \left( \mathbf{\Omega} \cross \mathbf{r} \right) \cross \hat{\mathbf{B}}$, would in principle shift the energies of most electronic states through changes in the spin-orbit and Darwin interactions, but this effect is likely far too small to observe with, e.g., optical lattice clocks.
We also note that, in the case of multiple closely-spaced atoms, the $[\hat{\bm{\mathcal{P}}}_{\perp}]^2$ contains cross-terms between different atoms that can lead to dipole-dipole interactions and thus must be retained explicitly in the Hamiltonian~\cite{Craig,Todorov}, but we will not consider this effect here. 

Finally, we discuss the effects of the intrinsic magnetic moment from Eq.~\eqref{H_MM}.
Above, we assumed that $\abs{\mathbf{B}} \sim \abs{\mathbf{E}} / c$ as is typical for the E\&M waves inside the cavity.
However, to distinguish the axis perpendicular to the plane of the cavity (which we call the $z$-axis with unit vector $\mathbf{z}$, see Fig.~\ref{Schematic}) as the quantization axis for the hyperfine splitting, it is common to introduce some weak external bias magnetic field along the $z$-axis~\cite{Derevianko2}.
We thus assume the magnetic field can be decomposed as,
\begin{equation}
    \hat{\mathbf{B}}_{\mathrm{tot}} = \hat{\mathbf{B}} + \mathbf{B}_{\mathrm{ext}},
\end{equation}
and so we get a term with $\hat{\mathbf{B}}$ and with $\mathbf{B}_{\mathrm{ext}}$ for each term that includes the magnetic field in Eq.~\eqref{H_H}.
We then can ignore the intrinsic magnetic moment's interaction with the quantized magnetic field as it is small compared to the electric dipole interaction,
\begin{equation}
    \begin{aligned}
& g_e e / (2 m_e) \abs{\mathbf{S}} \abs{\mathbf{B}} / (\abs{\mathbf{d}} \abs{\mathbf{E}}) \sim \hbar / (m_e c \text{\AA}) \ll 1, \\
& g_p e / (2 m_p) \abs{\mathbf{I}} \abs{\mathbf{B}} / (\abs{\mathbf{d}} \abs{\mathbf{E}}) \ll 1,
    \end{aligned}
\end{equation}
where we have used $\abs{\mathbf{S}} \sim \hbar$, $\abs{\mathbf{I}} \sim \hbar$, and $g_e e/(2 m_e) \gg g_p e/(2 m_p)$. 
Meanwhile, the intrinsic magnetic moment and the magnetic dipole interactions with the external field lead to the typical Hamiltonian for the Zeeman effect~\cite{Sakurai},
\begin{equation}
    \hat{H}_{\mathrm{Z}} = \left[ \frac{\mu_B}{\hbar} \left( \hat{\mathbf{L}} + g_e \hat{\mathbf{S}} \right) - \frac{g_p e}{2 m_p} \hat{\mathbf{I}} \right] \bm{\cdot} \mathbf{B}_{\mathrm{ext}} \equiv - \hat{\bm{\mu}} \bm{\cdot} \mathbf{B}_{\mathrm{ext}},
\end{equation}
with the Bohr magneton $\mu_B = e \hbar / (2 m_e)$.
The diamagnetic and R\"ontgen interactions involving the external magnetic field lead to small corrections to the Zeeman effect that are typically disregarded. 
We can then include the shift from the external field into the renormalized atomic internal Hamiltonian,
\begin{equation}
    \hat{H}'_{\mathrm{AI}} = \hat{H}_{\mathrm{AI}} + \hat{H}_{\mathrm{Z}} + \frac{1}{2 \epsilon_0} \int d^3 r \left[ \hat{\bm{\mathcal{P}}}_{\perp} (\mathbf{r}) \right]^2 + \tilde{\hat{H}}_{\mathrm{H,r}},
\end{equation}
such that the Zeeman effect from the bias field is again already incorperated into the physical atomic spectrum. 
Therefore, the atom-cavity Hamiltonian Eq.~\eqref{H_H} under the electric dipole approximation can be simplified to
\begin{equation} \label{H_simp}
    \begin{aligned}
\tilde{\hat{H}}_{\mathrm{H}} \approx & \frac{\hat{\mathbf{P}}^2}{2 M} + \hat{H}'_{\mathrm{AI}} + \frac{\epsilon_0}{2} \int d^3 r \left( \hat{E}_{\mathrm{F}}^2 + c^2 \hat{B}^2 \right) \\
& - \hat{\mathbf{d}} \bm{\cdot} \hat{\mathbf{E}}_{\mathrm{F}} (\hat{\mathbf{R}}) - \mathbf{\Omega} \bm{\cdot} \left( \hat{\mathbf{F}} + \hat{\bm{\mathfrak{L}}} \right) - \int d^3 r \mathbf{\Omega} \bm{\cdot} \hat{\bm{\ell}}_{\mathrm{F}}.
    \end{aligned}
\end{equation}

\subsection{Simplification of Quantized Electromagnetic Field} \label{Sec:EMsimplify}
The next step is to insert expressions for the quantized E\&M fields into our Hamiltonian. 
In Appendix~\ref{Appendix:EMquantization}, we quantize the vector potential $\hat{\mathbf{A}}$ by finding solutions of the Euler-Lagrange equation for the field Lagrangian in the Lorenz gauge and then transforming to the Coulomb gauge; we summarize the results below.
Note that we assume the cavity is empty throughout this subsection.
The periodic boundary conditions of the ring cavity give a discrete set of allowed frequencies when $\Omega = 0$,
\begin{equation} \label{omega_nI}
    \omega_n^I = c \abs{\mathbf{k}_n} = \frac{2 \pi c \abs{n}}{L_{\mathrm{path}}},
\end{equation}
where $n \in \{ \pm 1, \pm 2, \ldots \}$ and $L_{\mathrm{path}}$ is the path length of the cavity.
For each $\mathbf{k}_n$ and for different polarizations $\zeta$, we can introduce a creation (annihilation) operator $\hat{a}_{\mathbf{k}_n,\zeta}^{\dagger}$ ($\hat{a}_{\mathbf{k}_n,\zeta}$) such that the four-potential in the Lorenz gauge can be written in the general form~\cite{Birrell,Tong_QFT,Greiner}
\begin{equation}
    \hat{A}_{\mu} = i \sum_{\mathbf{k}_n} \sum_{\zeta = 0}^3 \epsilon_{\mu}^{\zeta} (\mathbf{r}, \mathbf{k}_n) \left[ f_{\mathbf{k}_n}(\mathbf{r},t) \hat{a}_{\mathbf{k}_n, \zeta} - f_{\mathbf{k}_n}^* (\mathbf{r},t) \hat{a}_{\mathbf{k}_n, \zeta}^{\dagger} \right].
\end{equation}
Since the spacetime is stationary, it possesses a timelike Killing vector [Eq.~\eqref{KillingMag}] which allows the mode functions to be separated into temporal and spatial components~\cite{Birrell,Wald}
\begin{equation}
    f_{\mathbf{k}_n} \propto e^{- i \omega_n t} u_{\mathbf{k}_n} (\mathbf{r}),
\end{equation}
for frequencies $\omega_n$ in the rotating frame. 
However, for a general rotation vector $\mathbf{\Omega}$, we cannot use separation of variables on the spatial degrees of freedom along the plane of the ring cavity. 
Therefore, we now make some simplifying assumptions about the system:
\begin{enumerate}[label=\roman*.]
    \item the polarization vectors are independent of position $\epsilon_{\mu}^{\zeta} (\mathbf{r}, \mathbf{k}_n) \approx \epsilon_{\mu}^{\zeta} (\mathbf{k}_n)$;
    
    \item we ignore the component of $\mathbf{\Omega}$ parallel to the plane of the ring cavity, i.e., we take $\mathbf{\Omega} \approx \tilde{\Omega} \mathbf{z} = \Omega \cos \theta \mathbf{z}$ where $\theta$ is the angle between $\mathbf{\Omega}$ and the $z$-axis (see Fig.~\ref{Schematic});

    \item the ring cavity is approximately a circular cavity of radius $\mathcal{R}$ (see Fig.~\ref{Schematic});

    \item and the field is confined in the radial and $z$ directions such that the ring cavity can be treated as a ring of radius $\mathcal{R}$ (see Fig.~\ref{Schematic}).
\end{enumerate}
With these simplifications [and now working in the Born metric Eq.~\eqref{g_B}], we can use separation of variables to write the four-potential in the one-dimensional ring as
\begin{equation}
    \begin{aligned}
\hat{A}_{\mu} = \sum_n & \sum_{\zeta = 0}^3 \sqrt{\frac{\hbar}{2 \epsilon_0 \omega_n^I \mathcal{V}}} \epsilon_{\mu}^{\zeta} (\mathbf{k}_n) \cross \\
& \left[ i e^{- i \left( \omega_n^I - k_n \tilde{\Omega} \mathcal{R} \right) t} e^{i k_n \mathcal{R} \phi} \hat{a}_{\mathbf{k}_n, \zeta} + \mathrm{H.c.} \right],
    \end{aligned}
\end{equation}
where $\mathcal{V}$ is the quantization volume and a modified dispersion relation gives the frequencies in the rotating frame,
\begin{equation}
    \omega_n = \omega_n^I - k_n \mathcal{R} \tilde{\Omega}.
\end{equation}
Finally, transforming from the Lorenz gauge to the Coulomb gauge removes the timelike and longitudinal polarizations with the result [Eq.~\eqref{A_tot_app}]
\begin{equation} \label{A_tot}
    \begin{aligned}
\hat{\mathbf{A}} = \sum_n & \sum_{\zeta = 1}^2 \sqrt{\frac{\hbar}{2 \epsilon_0 \omega_n^I \mathcal{V}}} \bm{\epsilon}_{\zeta} (\mathbf{k}_n) \cross \\
& \left[ i e^{- i \left( \omega_n^I - k_n \tilde{\Omega} \mathcal{R} \right) t} e^{i k_n \mathcal{R} \phi} \hat{a}_{\mathbf{k}_n, \zeta} + \mathrm{H.c.} \right],
    \end{aligned}
\end{equation}
where the polarization vectors of each mode are transverse to the propagation direction $\mathbf{k}_n \bm{\cdot} \bm{\epsilon}_{\zeta} (\mathbf{k}_n) = 0$.

\subsection{Two-mode, two-level, and rotating-wave approximations} \label{Sec:Approxs}
We can immediately plug Eq.~\eqref{A_tot} into the total Hamiltonian Eq.~\eqref{H_simp}, but it is common to make one further simplification of the E\&M field.
Here, one assumes that there is only one relevant frequency from Eq.~\eqref{omega_nI}, which we label $\omega_c$, in the atom-field interaction of interest. 
We can then drop every mode of the cavity except the modes with frequencies $\omega_{\pm} = \omega_c \mp \abs{k} \tilde{\Omega} \mathcal{R}$ in the co-rotating ($\omega_+$) and counter-rotating ($\omega_-$) directions, where we have dropped the subscript on the wave vector amplitude.
Finally, we assume there is only one relevant polarization for each mode $\bm{\epsilon}_{\pm}$ and then label the relevant co- and counter-rotating modes as $\hat{a}_{\pm}$.
This amounts to a \emph{two-mode} approximation (i.e., the typical single-mode approximation in both directions) in which the vector potential becomes [Eq.~\eqref{A_tm_app}]
\begin{equation}
    \begin{aligned}
\hat{\mathbf{A}} &= \sqrt{\frac{\hbar}{2 \epsilon_0 \omega_c \mathcal{V}}} \bm{\epsilon}_+ \left[ i e^{- i \left( \omega_c - \abs{k} \tilde{\Omega} \mathcal{R} \right) t} e^{i \abs{k} \mathcal{R} \phi} \hat{a}_+ + \mathrm{H.c.} \right] \\
& + \sqrt{\frac{\hbar}{2 \epsilon_0 \omega_c \mathcal{V}}} \bm{\epsilon_-} \left[ i e^{- i \left( \omega_c + \abs{k} \tilde{\Omega} \mathcal{R} \right) t} e^{- i \abs{k} \mathcal{R} \phi} \hat{a}_- + \mathrm{H.c.} \right],
    \end{aligned}
\end{equation}
from which we get the physical free fields for the rotating observer [Eq.~\eqref{E_quant_final}],
\begin{equation}
    \begin{aligned}
\hat{\mathbf{E}}_{\mathrm{F}} = & - \sqrt{\frac{\hbar \omega_c}{2 \epsilon_0 \mathcal{V}}} \bm{\epsilon}_+ \left[ e^{- i \left( \omega_c - \abs{k} \tilde{\Omega} \mathcal{R} \right) t} e^{i \abs{k} \mathcal{R} \phi} \hat{a}_+ + \mathrm{H.c.} \right] \\
& - \sqrt{\frac{\hbar \omega_c}{2 \epsilon_0 \mathcal{V}}} \bm{\epsilon_-} \left[ e^{- i \left( \omega_c + \abs{k} \tilde{\Omega} \mathcal{R} \right) t} e^{- i \abs{k} \mathcal{R} \phi} \hat{a}_- + \mathrm{H.c.} \right],
    \end{aligned}
\end{equation}
and [Eq.~\eqref{B_quant}]
\begin{equation}
    \begin{aligned}
\hat{\mathbf{B}} &= - \sqrt{\frac{\hbar \omega_c}{2 \epsilon_0 \mathcal{V}}} \frac{\bm{\phi} \cross \bm{\epsilon}_+}{c} \left[ e^{- i \left( \omega_c - \abs{k} \tilde{\Omega} \mathcal{R} \right) t} e^{i \abs{k} \mathcal{R} \phi} \hat{a}_+ + \mathrm{H.c.} \right] \\
& + \sqrt{\frac{\hbar \omega_c}{2 \epsilon_0 \mathcal{V}}} \frac{\bm{\phi} \cross \bm{\epsilon}_-}{c} \left[ e^{- i \left( \omega_c + \abs{k} \tilde{\Omega} \mathcal{R} \right) t} e^{- i \abs{k} \mathcal{R} \phi} \hat{a}_- + \mathrm{H.c.} \right].
    \end{aligned}
\end{equation}
Upon plugging in the quantized E\&M fields, the total free field Hamiltonian Eq.~\eqref{H_F} can be written as [Eq.~\eqref{H_F_app}]
\begin{equation}
    \begin{aligned}
\hat{H}_{\mathrm{F}} &= :\hat{H}_{\mathrm{IF}}: + :\hat{H}_{\mathrm{SF}}: \\
&= \hbar \left( \omega_c - \abs{k} \tilde{\Omega} \mathcal{R} \right) \hat{a}_+^{\dagger} \hat{a}_+ + \hbar \left( \omega_c + \abs{k} \tilde{\Omega} \mathcal{R} \right) \hat{a}_-^{\dagger} \hat{a}_-,
    \end{aligned}
\end{equation}
where we have applied normal ordering $: \hat{a}_{\pm} \hat{a}_{\pm}^{\dagger} : = \hat{a}_{\pm}^{\dagger} \hat{a}_{\pm}$ and $: \hat{a}_{\pm}^{\dagger} \hat{a}_{\pm} : = \hat{a}_{\pm}^{\dagger} \hat{a}_{\pm}$ to avoid the vacuum energy.
Thus, we see that the co-rotating mode is redshifted with respect to its inertial frequency while the counter-rotating mode is blueshifted, as expected from the Sagnac effect~\cite{Post}.

With this, we can then make the \emph{two-level} approximation on the atom in which the $\hat{a}_{\pm}$ modes are taken to be near-resonant with a specific transition (e.g., the hydrogen-alpha line~\cite{Sakurai}) such that the population in all the other energy levels can be neglected~\cite{Meystre,Berman,Frasca}.
In general, this can instead be a few-level approximation where all possible transitions between the angular momentum multiplets of the two energy levels are considered or there are multiple light fields each near-resonant with a few different transitions~\cite{CohenTannoudji2,Metcalf,Evers}, but we will not consider this here. 
Therefore, we assume the atom has an electronic ground state $\ket{\downarrow}$ and excited state $\ket{\uparrow}$ separated in the non-rotating frame by an optical transition frequency $\omega_a$. 
The modified internal Hamiltonian of the atom is then approximated as~\cite{Mandel}
\begin{equation}
    \hat{H}'_{\mathrm{AI}} \approx \left( E_a + \frac{\hbar \omega_a}{2} \right) \op{\uparrow}{\uparrow} + \left( E_a - \frac{\hbar \omega_a}{2} \right) \op{\downarrow}{\downarrow},
\end{equation}
where $E_a$ is the energy half-way between the two energy levels. 
We can then include the rotation-induced hyperfine shift Eq.~\eqref{HyperfineShift} by assuming that the ground and excited states are in the magnetic sublevels $\ket{\downarrow} = \ket{F,m_F}$ and $\ket{\uparrow} = \ket{F',m_{F'}}$.
Thus, the states are shifted by
\begin{equation} \label{AIpRH}
    \begin{aligned}
\hat{H}'_{\mathrm{AI}} + \hat{H}_{\mathrm{RH}} &\approx \left( E_a + \frac{\hbar \omega_a}{2} - \hbar \tilde{\Omega} m_{F'} \right) \op{\uparrow}{\uparrow} \\
& + \left( E_a - \frac{\hbar \omega_a}{2} - \hbar \tilde{\Omega} m_F \right) \op{\downarrow}{\downarrow}.
    \end{aligned}
\end{equation}
Here, we are assuming the external magnetic field $\mathbf{B}_{\mathrm{ext}}$ determines the quantization axis for hyperfine splitting rather than the rotation which is usually a good assumption; e.g., for $\abs{\mathbf{B}_{\mathrm{ext}}} \sim \mathrm{G}$ and $\abs{\Omega} \sim \mathrm{Hz}$, we estimate $\mu_B \abs{\mathbf{B}_{\mathrm{ext}}} / (\abs{\Omega} \abs{\mathbf{F}}) \sim 10^7$ where we have used $\abs{\mathbf{F}} \sim \hbar$ and $\mu_B \sim 10^{-23} \; \mathrm{J}/\mathrm{T}$. 
We then move into an interaction picture defined by the unitary operator, 
\begin{equation}
    \hat{U} = \mathrm{exp} \left[ - i t \left( \frac{E_a}{\hbar} - \frac{\tilde{\Omega} [m_{F'} + m_F]}{2} \right) (\op{\uparrow}{\uparrow} + \op{\downarrow}{\downarrow}) \right],
\end{equation}
which in the two-level approximation is proportional to an identity operator $\hat{\mathbb{I}} \approx \op{\uparrow}{\uparrow} + \op{\downarrow}{\downarrow}$ such that the unitary simply corresponds to redefining the point of zero energy in the atom,
\begin{equation}
    \tilde{\hat{H}}'_{\mathrm{AI}} + \tilde{\hat{H}}_{\mathrm{RH}} \approx \frac{\hbar}{2} \left( \omega_a - \tilde{\Omega} \Delta m \right) \left( \op{\uparrow}{\uparrow} - \op{\downarrow}{\downarrow} \right),
\end{equation}
where $\Delta m = m_{F'} - m_F$. 

Finally, we examine the electric dipole interaction. 
Within the two-level approximation, the dipole operator can be written in terms of off-diagonal elements of the bare states,
\begin{equation}
    \hat{\mathbf{d}} \approx \bra{\uparrow} \hat{\mathbf{d}} \ket{\downarrow} \op{\uparrow}{\downarrow} + (\bra{\uparrow} \hat{\mathbf{d}} \ket{\downarrow})^* \op{\downarrow}{\uparrow} = \hat{\mathbf{d}}_{\parallel} \hat{\sigma}_+ + \hat{\mathbf{d}}_{\parallel}^* \hat{\sigma}_-,
\end{equation}
using parity arguments~\cite{Mandel}.
Here, we defined $\hat{\mathbf{d}}_{\parallel} \equiv \bra{\uparrow} \hat{\mathbf{d}} \ket{\downarrow}$ and the Pauli operators (distinct from those for quantum spin in Sec.~\ref{Sec:DiracEqCurvedST}) $\hat{\sigma}_+ = \op{\uparrow}{\downarrow} = \hat{\sigma}_-^{\dagger}$ and
\begin{equation}
    \hat{\sigma}_x = \hat{\sigma}_+ + \hat{\sigma}_-, \quad \hat{\sigma}_y = i \left( \hat{\sigma}_- - \hat{\sigma}_+ \right), \quad \hat{\sigma}_z = \left[ \hat{\sigma}_+, \hat{\sigma}_- \right].
\end{equation}
The electric dipole interaction then becomes
\begin{equation}
    \begin{aligned}
\hat{H}_{\mathrm{AE}} = & - \hat{\mathbf{d}} \bm{\cdot} \hat{\mathbf{E}}_{\mathrm{F}} (\hat{\mathbf{R}}) \\
= & \sqrt{\frac{\hbar \omega_c}{2 \epsilon_0 \mathcal{V}}} \left( \hat{\mathbf{d}}_{\parallel} \hat{\sigma}_+ + \mathrm{H.c.} \right) \bm{\cdot} \left( \bm{\epsilon}_+ \left[ e^{i \abs{k} \mathcal{R} \hat{\phi}} \hat{a}_+ + \mathrm{H.c.} \right] \right. \\
& \left. + \bm{\epsilon_-} \left[ e^{- i \abs{k} \mathcal{R} \hat{\phi}} \hat{a}_- + \mathrm{H.c.} \right] \right) \\
= & \hbar \left( g_+ \hat{\sigma}_+ + g_+^* \hat{\sigma}_- \right) \left( e^{i \abs{k} \mathcal{R} \hat{\phi}} \hat{a}_+ + e^{- i \abs{k} \mathcal{R} \hat{\phi}} \hat{a}_+^{\dagger} \right) \\
& + \hbar \left( g_- \hat{\sigma}_+ + g_-^* \hat{\sigma}_- \right) \left( e^{- i \abs{k} \mathcal{R} \hat{\phi}} \hat{a}_- + e^{i \abs{k} \mathcal{R} \hat{\phi}} \hat{a}_-^{\dagger} \right),
    \end{aligned}
\end{equation}
with the atom-cavity coupling constants,
\begin{equation}
    g_{\pm} = \sqrt{\frac{\omega_c}{2 \hbar \epsilon_0 \mathcal{V}}} \hat{\mathbf{d}}_{\parallel} \bm{\cdot} \bm{\epsilon}_{\pm}.
\end{equation}
The full Hamiltonian Eq.~\eqref{H_simp} is then given by
\begin{equation}
    \begin{aligned}
\tilde{\hat{H}}_{\mathrm{H}} = & \hbar \omega_c \left( \hat{a}_+^{\dagger} \hat{a}_+ + \hat{a}_-^{\dagger} \hat{a}_- \right) - \hbar \abs{k} \tilde{\Omega} \mathcal{R} \left( \hat{a}_+^{\dagger} \hat{a}_+ - \hat{a}_-^{\dagger} \hat{a}_- \right) \\
& + \frac{\hat{\mathbf{P}}^2}{2 M} - \tilde{\Omega} \mathbf{z} \bm{\cdot} \hat{\bm{\mathfrak{L}}} + \frac{\hbar}{2} \left( \omega_a - \tilde{\Omega} \Delta m \right) \hat{\sigma}_z \\
& + \hbar \left( g_+ \hat{\sigma}_+ + g_+^* \hat{\sigma}_- \right) \left( e^{i \abs{k} \mathcal{R} \hat{\phi}} \hat{a}_+ + e^{- i \abs{k} \mathcal{R} \hat{\phi}} \hat{a}_+^{\dagger} \right) \\
& + \hbar \left( g_- \hat{\sigma}_+ + g_-^* \hat{\sigma}_- \right) \left( e^{- i \abs{k} \mathcal{R} \hat{\phi}} \hat{a}_- + e^{i \abs{k} \mathcal{R} \hat{\phi}} \hat{a}_-^{\dagger} \right).
    \end{aligned}
\end{equation}

Next, we make the \emph{rotating-wave} approximation (RWA).
Here, we transform into an interaction picture defined by
\begin{equation} \label{U_RWA}
    \begin{aligned}
& \hat{U}_{\mathrm{RWA}} = \exp \left[ - \frac{i t}{2} (\omega_a - \tilde{\Omega} \Delta m) \hat{\sigma}_z \right] \cross \\
& \exp \left[ - i t \left( \omega_c - \abs{k} \tilde{\Omega} \mathcal{R} \right) \hat{a}_+^{\dagger} \hat{a}_+ - i t \left( \omega_c + \abs{k} \tilde{\Omega} \mathcal{R} \right) \hat{a}_-^{\dagger} \hat{a}_- \right],
    \end{aligned}
\end{equation}
which rotates the interaction terms by
\begin{equation} \label{RWA}
    \begin{aligned}
\hat{\sigma}_+ \hat{a}_{\pm} &\longrightarrow \hat{\sigma}_+ \hat{a}_{\pm} e^{i (\omega_a - \tilde{\Omega} \Delta m) t - i (\omega_c \mp \abs{k} \tilde{\Omega} \mathcal{R}) t} \\ 
\hat{\sigma}_- \hat{a}_{\pm}^{\dagger} &\longrightarrow \hat{\sigma}_- \hat{a}_{\pm}^{\dagger} e^{- i (\omega_a - \tilde{\Omega} \Delta m) t + i (\omega_c \mp \abs{k} \tilde{\Omega} \mathcal{R}) t} \\
\hat{\sigma}_+ \hat{a}_{\pm}^{\dagger} &\longrightarrow \hat{\sigma}_+ \hat{a}_{\pm}^{\dagger} e^{i (\omega_a - \tilde{\Omega} \Delta m) t + i (\omega_c \mp \abs{k} \tilde{\Omega} \mathcal{R}) t} \\ 
\hat{\sigma}_- \hat{a}_{\pm} &\longrightarrow \hat{\sigma}_- \hat{a}_{\pm} e^{- i (\omega_a - \tilde{\Omega} \Delta m) t - i (\omega_c \mp \abs{k} \tilde{\Omega} \mathcal{R}) t}.
    \end{aligned}
\end{equation}
Since $\abs{(\omega_a - \tilde{\Omega} \Delta m) + (\omega_c \mp \abs{k} \tilde{\Omega} \mathcal{R})}$ are optical frequencies $\mathcal{O}(100 \, \mathrm{THz})$ while $\abs{(\omega_a - \tilde{\Omega} \Delta m) - (\omega_c \mp \abs{k} \tilde{\Omega} \mathcal{R})}$ are assumed to be a microwave or radio wave frequencies $\mathcal{O}(100 \; \mathrm{GHz})$ or smaller, we can make a coarse-graining approximation for the fast oscillating terms which averages the final two terms in Eq.~\eqref{RWA} to zero over the timescale of interest~\cite{Meystre,Steck,Mandel}.
Moving back out of the rotating frame from Eq.~\eqref{U_RWA}, we at last obtain a modified Jaynes-Cummings model in the rotating cavity,
\begin{widetext}
\begin{equation} \label{H_final}
    \begin{aligned}
\tilde{\hat{H}}_{\mathrm{H}} = & \frac{\hat{\mathbf{P}}^2}{2 M} - \tilde{\Omega} \mathbf{z} \bm{\cdot} \hat{\bm{\mathfrak{L}}} + \frac{\hbar}{2} \left( \omega_a - \tilde{\Omega} \Delta m \right) \hat{\sigma}_z + \hbar \omega_c \left( \hat{a}_+^{\dagger} \hat{a}_+ + \hat{a}_-^{\dagger} \hat{a}_- \right) - \hbar \abs{k} \tilde{\Omega} \mathcal{R} \left( \hat{a}_+^{\dagger} \hat{a}_+ - \hat{a}_-^{\dagger} \hat{a}_- \right) \\
& + \hbar \abs{g_+} \left[ e^{i \vartheta_+} e^{i \abs{k} \mathcal{R} \hat{\phi}} \hat{\sigma}_+ \hat{a}_+ + e^{- i \vartheta_+} e^{- i \abs{k} \mathcal{R} \hat{\phi}} \hat{a}_+^{\dagger} \hat{\sigma}_- \right] + \hbar \abs{g_-} \left[ e^{i \vartheta_-} e^{- i \abs{k} \mathcal{R} \hat{\phi}} \hat{\sigma}_+ \hat{a}_- + e^{- i \vartheta_-} e^{i \abs{k} \mathcal{R} \hat{\phi}} \hat{a}_-^{\dagger} \hat{\sigma}_- \right],
    \end{aligned}
\end{equation}
\end{widetext}
with $\vartheta_{\pm} = \mathrm{arg}(g_{\pm})$.

As a final step, we can extend the system to the case of $N$ protium atoms inside the cavity.
As discussed around Eq.~\eqref{H_AIp}, the $[\hat{\bm{\mathcal{P}}}_{\perp}]^2$ term can lead to dipole-dipole interactions between closely-spaced atoms, but we will only consider non-interacting particles here. 
We take atom $j$ to be at position $\hat{\phi}_j$ in the ring and to have the non-relativistic momentum $\hat{\mathbf{P}}_j$, as well as angular momentum $\hat{\bm{\mathfrak{L}}}_j$. 
Atom $j$ then interacts with the two modes of the cavity with a coupling constant,
\begin{equation}
    g_j^{\pm} = \sqrt{\frac{\omega_c}{2 \hbar \epsilon_0 \mathcal{V}}} \hat{\mathbf{d}}_j^{\parallel} \bm{\cdot} \mathbf{\epsilon}_{\pm},
\end{equation}
with the dipole component $\hat{\mathbf{d}}_j^{\parallel} = \bra{\uparrow}_j \hat{\mathbf{d}}_j \ket{\downarrow}_j$. 
After the RWA, the electric dipole interaction for atom $j$ becomes
\begin{equation}
    \begin{aligned}
\hat{H}_{\mathrm{AE}}^{(j)} &= \hbar \abs{g_j^+} \left( e^{i \vartheta_j^+} e^{i \abs{k} \mathcal{R} \hat{\phi}_j} \hat{\sigma}_+^{(j)} \hat{a}_+ + e^{- i \vartheta_j^+} e^{- i \abs{k} \mathcal{R} \hat{\phi}_j} \hat{a}_+^{\dagger} \hat{\sigma}_-^{(j)} \right) \\
& + \hbar \abs{g_j^-} \left( e^{i \vartheta_j^-} e^{- i \abs{k} \mathcal{R} \hat{\phi}_j} \hat{\sigma}_+^{(j)} \hat{a}_- + e^{- i \vartheta_j^-} e^{i \abs{k} \mathcal{R} \hat{\phi}_j} \hat{a}_-^{\dagger} \hat{\sigma}_-^{(j)} \right),
    \end{aligned}
\end{equation}
where $\vartheta_j^{\pm} = \mathrm{arg}(g_j^{\pm})$ and $\hat{\sigma}_+^{(j)} = \op{\uparrow}{\downarrow}_j = \left( \hat{\sigma}_-^{(j)} \right)^{\dagger}$ are the Pauli operators for atom $j$.
With this, we obtain a modified Tavis-Cummings Hamiltonian in the rotating frame,
\begin{widetext}
\begin{equation}
    \begin{aligned}
& \tilde{\hat{H}}_{\mathrm{H}, N} = \hbar \omega_c \left( \hat{a}_+^{\dagger} \hat{a}_+ + \hat{a}_-^{\dagger} \hat{a}_- \right) - \hbar \abs{k} \tilde{\Omega} \mathcal{R} \left( \hat{a}_+^{\dagger} \hat{a}_+ - \hat{a}_-^{\dagger} \hat{a}_- \right) + \sum_{j = 1}^N \left[ \frac{\hat{\mathbf{P}}_j^2}{2 M} - \tilde{\Omega} \mathbf{z} \bm{\cdot} \hat{\bm{\mathfrak{L}}}_j + \frac{\hbar}{2} \left( \omega_a - \tilde{\Omega} \Delta m \right) \hat{\sigma}_z^{(j)} \right] \\
& + \hbar \sum_{j = 1}^N \left( \abs{g_j^+} \left[ e^{i \vartheta_j^+} e^{i \abs{k} \mathcal{R} \hat{\phi}_j} \hat{\sigma}_+^{(j)} \hat{a}_+ + e^{- i \vartheta_j^+} e^{- i \abs{k} \mathcal{R} \hat{\phi}_j} \hat{a}_+^{\dagger} \hat{\sigma}_-^{(j)} \right] + \abs{g_j^-} \left[ e^{i \vartheta_j^-} e^{- i \abs{k} \mathcal{R} \hat{\phi}_j} \hat{\sigma}_+^{(j)} \hat{a}_- + e^{- i \vartheta_j^-} e^{i \abs{k} \mathcal{R} \hat{\phi}_j} \hat{a}_-^{\dagger} \hat{\sigma}_-^{(j)} \right] \right).
    \end{aligned}
\end{equation}
\end{widetext}

\section{Conclusion and Outlook} \label{Sec:Conclusion}
In this paper, we have derived a modified Jaynes-Cummings model in a rotating ring cavity from first principles: the Dirac equation in curved spacetime. 
After taking the non-relativistic limit of a protium atom, a PWZ transformation reveals a rotation-induced hyperfine shift Eq.~\eqref{HyperfineShift} and a rotation-induced R\"ontgen interaction~\eqref{RotationRontgen}.
We again emphasize the requirement to begin from a full, covariant field-theoretic approach when examining the effects of non-inertial spacetimes in quantum optics~\cite{Falcone}, as the two predicted shifts [Eqs.~\eqref{HyperfineShift} and~\eqref{RotationRontgen}] would not emerge naturally in an entirely non-relativistic approach and instead are added \emph{a posteriori}.
This point is even more crucial in more general spacetimes, especially spacetimes with non-zero curvature $R^{\rho}_{\phantom{\rho} \sigma \mu \nu} \neq 0$ where the non-relativistic limit (FW transformation) requires much more care due to consequences of the equivalence principle (namely, the gravitational ``charge'' being the same as the inertial mass about which we are expanding)~\cite{Buhl} and non-trivial gravitational couplings can arise~\cite{Falcone,Sorge,Mashhoon3,Obukhov2,Silenko4}. 
We have already touched upon the possibility of measuring the rotation-induced hyperfine shift with a superradiant laser~\cite{Meiser,Norcia,Reilly}, and one would expect the shift from Earth's rotation to potentially impact optical lattice clock measurements soon; for the $429.2 \, \mathrm{THz}$ clock transition in ${}^{87} \mathrm{Sr}$, a state-of-the-art fractional frequency uncertainty of $8.1 \cross 10^{-19}$~\cite{Aeppli} is at the $\mathcal{O}(300 \, \mu\mathrm{Hz})$-level while the rotation rate of the Earth is $11.6 \, \mu\mathrm{Hz}$ [the hyperfine shift has an additional factor from $m_F = \pm 5/2$ and $m_{F'} = \pm 3/2$, see Eq.~\eqref{AIpRH}].
On the other hand, the rotational R\"ontgen interaction is minuscule in typical cavity QED experiments, but may become relevant in systems with large magnetic fields.
For example, this shift may produce measurable corrections on spectroscopic measurements of atoms and molecules in extreme magnetic fields around astrophysical objects, such as black holes~\cite{Chatterjee,Roelofs,Akiyama} and magnetars~\cite{Turolla,Kaspi}.

The most immediate generalization of our derivation is to avoid the simplifications in Secs.~\ref{Sec:Approxs} and~\ref{Sec:SepVariables} that were required to use separation of variables on the spatial degrees of freedom of the E\&M mode functions.
Instead, one wishes to derive the quantized field Hamiltonian and electric and magnetic fields with the appropriate Sagnac shifts for a completely general cavity geometry and rotation axis. 
Moreover, it would be interesting to include the polarization-transport effects that were neglected after Eq.~\eqref{WaveEq_Pol} which can become relevant when the rotation axis has a large component along the plane of the cavity~\cite{Mashhoon6}, and effects of backscattering off the mirrors which couples the counterpropagating modes~\cite{Chow,Faucheux}; both of these effects can impact results in a ring laser gyroscope experiment. 
The next obvious generalization is to more complicated atomic species as rotation-induced shifts in alkali(-like) atoms could have an impact on fountain clock measurements~\cite{Inguscio}.
Moreover, as discussed before, these shifts are potentially even more important in alkaline-earth(-like) atoms due to the extreme precision of optical lattice atomic clocks~\cite{Derevianko,Ludlow,Hollberg,Fortier,Inguscio,Bothwell} and superradiant lasers~\cite{Meiser,Norcia,Reilly}. 
For example, a superradiant ring laser gyroscope may be able to directly measure $\Omega$ through simple heterodyne measurements of the laser's frequency as it is slaved to the atomic resonance frequency~\cite{Reilly2}.
Another intriguing possibility is that spin-rotation coupling could potentially provide an alternative mechanism to the quadratic Zeeman shift from a magnetic field in coupling the ${}^3 P_0$ clock state to ${}^3 P_1$ in bosonic alkaline-earth isotopes (e.g., ${}^{88} \mathrm{Sr}$~\cite{Taichenachev,Lu,Dong}).
While the first-order shift from Eq.~\eqref{HyperfineShift} does not couple the fine-split states as $\bra{{}^3 P_1} \mathbf{\Omega} \bm{\cdot} \hat{\mathbf{J}} \ket{{}^3 P_0} = 0$ (with $I = 0$) since it cannot couple inequivalent $\mathrm{SU}(2)$ irreducible representations (generated by $\hat{\mathbf{J}}$) by Schur’s lemma~\cite{Georgi}, it is possible that higher-order spin-coupling terms in the FW transformation or other non-inertial spin-coupling effects~\cite{Hehl} could potentially couple these states as $\bra{{}^3 P_1} \hat{\mathbf{S}} \ket{{}^3 P_0} \neq 0$.
Generalizing our calculation to these more complicated atoms typically entails considering the nucleus and inner shell electrons as an ionic core fixed at the CoM position and then adding the valence electrons in a systematic way~\cite{Santra}, while relativistic corrections can be estimated by a Breit-like approach~\cite{Breit,Bethe}.
Finally, it would be interesting to extend our calculation to an open quantum system coupled to free-space electromagnetic modes, where spontaneous emission or cavity decay may potentially contain imprints of inertial effects~\cite{Yu,Paczos}.

\section*{Acknowledgments}
We would like to thank Charles Marrder, John Cooper, Simon B. J\"ager, John D. Wilson, Haonan Liu, and Andrew J. S. Hamilton for useful discussions. 
This material is based upon work supported by the National Science Foundation Grant Nos.\ 2016244 and 2317149.

\appendix

\section{Details of Foldy-Wouthuysen Transformation} \label{Appendix:FWtransform}
This appendix is dedicated to evaluating the Hamiltonian to order $\mathcal{O} (1 / m^2)$ after the FW transformation, which is given in Eq.~\eqref{FWtransform}.
Note that we use natural units throughout this appendix.
We begin with the even and odd components of the original Hamiltonian, given in
Eq.~\eqref{EvenOdd}:
\begin{equation}
    \begin{aligned}
\mathcal{E} &= q \varphi - \mathbf{\Omega} \bm{\cdot} \mathbf{J} + q \mathbf{\Omega} \bm{\cdot} \left( \mathbf{r} \cross \mathbf{A} \right), \\ 
\mathcal{O} &= \bm{\alpha} \bm{\cdot} (\mathbf{p} - q \mathbf{A}),
    \end{aligned}
\end{equation}
The $\beta m$ and $\mathcal{E}$ terms in Eq.~\eqref{FWtransform} are trivial, and so to obtain the first line of Eq.~\eqref{FWtransform}, we only need to work out $\mathcal{O}^2$ and $\mathcal{O}^4$.
Using $\sigma_i \sigma_j = \delta_{ij} + i \varepsilon_{ijk} \sigma_k$, we then find
\begin{equation}
    \mathcal{O}^2 = (\mathbf{p} - q \mathbf{A})^2 - 2 q \mathbf{S} \bm{\cdot} \mathbf{B},
\end{equation}
from which we obtain
\begin{equation}
    \frac{\mathcal{O}^4}{8 m^3} = \frac{(\mathbf{p} - q \mathbf{A})^4}{8 m^3} - \frac{q}{4 m^3} \left\{ (\mathbf{p} - q \mathbf{A})^2, \mathbf{S} \bm{\cdot} \mathbf{B} \right\} + \frac{q^2}{2 m^3} \left( \mathbf{S} \bm{\cdot} \mathbf{B} \right)^2.
\end{equation}
In the weak field limit, the $\mathbf{p}^4$ term is the only term of the desired order $\mathcal{O} (1 / m^2)$~\cite{Bjorken}, and so we take
\begin{equation}
    \frac{\mathcal{O}^4}{8 m^3} \approx \frac{\mathbf{p}^4}{8 m^3}.
\end{equation}
Therefore, the term in the parenthesis on the first line of Eq.~\eqref{FWtransform} becomes
\begin{equation}
    m + \frac{\mathcal{O}^2}{2 m} - \frac{\mathcal{O}^4}{8 m^3} \approx m + \frac{(\mathbf{p} - q \mathbf{A})^2}{2 m} - \frac{q}{m} \mathbf{S} \bm{\cdot} \mathbf{B} - \frac{\mathbf{p}^4}{8 m^3},
\end{equation}
which has the Taylor expansion of the particle's energy, $\sqrt{m^2 + \mathbf{p}^2}$, to second order~\cite{Bjorken}.
The first line of Eq.~\eqref{FWtransform} is thus given by
\begin{equation} \label{FW_FirstLine}
    \begin{aligned}
\beta \left( m + \frac{\mathcal{O}^2}{2 m} \right. & \left. - \frac{\mathcal{O}^4}{8 m^3} \right) + \mathcal{E} = \beta \left[ m + \frac{(\mathbf{p} - q \mathbf{A})^2}{2 m} - \frac{\mathbf{p}^4}{8 m^3} \right] \\
& - \frac{q}{m} \beta \mathbf{S} \bm{\cdot} \mathbf{B} + q \varphi - \mathbf{\Omega} \bm{\cdot} \mathbf{J} + q \mathbf{\Omega} \bm{\cdot} \left( \mathbf{r} \cross \mathbf{A} \right).
    \end{aligned}
\end{equation}

We now work out the commutator on the second line of Eq.~\eqref{FWtransform}.
We first find
\begin{equation}
    \rpd{t} \mathcal{O} = - q \bm{\alpha} \bm{\cdot} \rpd{t} \mathbf{A},
\end{equation}
and the non-zero commutators in $[\mathcal{O}, \mathcal{E}]$,
\begin{equation}
    \begin{aligned}
\left[ \mathcal{O}, \varphi \right] &= - i \mathbf{\alpha} \bm{\cdot} \bm{\nabla} \varphi, \\
- \left[ \mathcal{O}, \mathbf{\Omega} \bm{\cdot} \mathbf{L} \right] &= i \bm{\alpha} \bm{\cdot} \left( \mathbf{p} \cross \mathbf{\Omega} \right) + i q \left[ \left( \mathbf{\Omega} \cross \mathbf{r} \right) \bm{\cdot} \bm{\nabla} \right] \left( \bm{\alpha} \bm{\cdot} \mathbf{A} \right), \\
- \left[ \mathcal{O}, \mathbf{\Omega} \bm{\cdot} \mathbf{S} \right] &= - i \bm{\alpha} \bm{\cdot} \left[ (\mathbf{p} - q \mathbf{A}) \cross \mathbf{\Omega} \right], \\
\left[ \mathcal{O}, \mathbf{\Omega} \bm{\cdot} \left( \mathbf{r} \cross \mathbf{A} \right) \right] &= i \bm{\alpha} \bm{\cdot} \left( \mathbf{\Omega} \cross \mathbf{A} \right) - i \left( \mathbf{\Omega} \cross \mathbf{r} \right) \bm{\cdot} \left[ (\bm{\alpha} \bm{\cdot} \bm{\nabla}) \mathbf{A} \right],
    \end{aligned}
\end{equation}
where one must remember that $\mathbf{S}$ has an implicit $\mathbb{I}$ with it when doing the commutator with $\bm{\alpha}$ and we have used (non-commutative) vector algebra identities. 
Using the definition of the electric field for the rotating observer Eq.~\eqref{EBdef}, we then obtain
\begin{widetext}
\begin{equation}
    \begin{aligned}
\left[ \mathcal{O}, \mathcal{E} \right] + i \rpd{t} \mathcal{O} = & - i q \bm{\alpha} \bm{\cdot} \bm{\nabla} \varphi - i \bm{\alpha} \bm{\cdot} \left( \mathbf{p} \cross \mathbf{\Omega} \right) + i q \bm{\alpha} \bm{\cdot} \left( \mathbf{A} \cross \mathbf{\Omega} \right) + i \bm{\alpha} \bm{\cdot} \left( \mathbf{p} \cross \mathbf{\Omega} \right) \\
& + i q \left[ \left( \mathbf{\Omega} \cross \mathbf{r} \right) \bm{\cdot} \bm{\nabla} \right] \left( \bm{\alpha} \bm{\cdot} \mathbf{A} \right) + i q \bm{\alpha} \bm{\cdot} \left( \mathbf{\Omega} \cross \mathbf{A} \right) - i q \left( \mathbf{\Omega} \cross \mathbf{r} \right) \bm{\cdot} \left[ (\bm{\alpha} \bm{\cdot} \bm{\nabla}) \mathbf{A} \right] - i q \bm{\alpha} \bm{\cdot} \rpd{t} \mathbf{A} \\
= & i q \bm{\alpha} \bm{\cdot} \mathbf{E},
    \end{aligned}
\end{equation}
\end{widetext}
where we have again used the three-dimensional Levi-Civita pseudotensor relation $\varepsilon_{ijk} \varepsilon^{mnk} = \delta_i^m \delta_j^n - \delta_i^n \delta_j^m$~\cite{Landau}.
We then find that the second line of Eq.~\eqref{FWtransform} is given by
\begin{equation} \label{FW_SecondLine}
    \begin{aligned}
\left[ \left( \left[ \mathcal{O}, \mathcal{E} \right] + i \rpd{t} \mathcal{O} \right), \frac{\mathcal{O}}{8 m^2} \right] = & - \frac{i q}{8 m^2} \Big( \mathbf{p} \bm{\cdot} \mathbf{E} \\
& + 2 i \mathbf{S} \bm{\cdot} \left[ (\mathbf{p} - q \mathbf{A}), \mathbf{E} \right]_{\cross} \Big),
    \end{aligned}
\end{equation}
where we have dropped an identity term from the $[\mathbf{A}, \mathbf{E}]$ commutator as it can be dropped in the final Hamiltonian.
Inserting Eqs.~\eqref{FW_FirstLine} and~\eqref{FW_SecondLine} back into Eq.~\eqref{FWtransform}, we arrive at the block-diagonal Hamiltonian to the desired order in Eq.~\eqref{H'_D}.

\section{Quantization of the Electromagnetic Field} \label{Appendix:EMquantization}
In this appendix, we quantize the E\&M field inside the rotating ring cavity. 
To accurately account for the time-space mixing from the $g_{0j}$ components of the metric tensor, it is convenient to first quantize the field in the covariant Lorenz gauge and then gauge transform to the Coulomb gauge whereupon it can be plugged into the Hamiltonian in Eq.~\eqref{H_F}.
We assume there are no atoms inside the cavity throughout this appendix. 

\subsection{Wave equation}
We begin with the electromagnetic field in the Lorenz gauge~\cite{Griffiths_EM},
\begin{equation}
    \rcd{\mu} A^{\mu} = 0.
\end{equation}
The Lagrangian can be written (in natural units) as~\cite{Tong_QFT,Tong_GT}
\begin{equation}
    \mathcal{L}_{\mathrm{F}} = - \frac{1}{4} F^{\mu \nu} F_{\mu \nu} - \frac{1}{2 \xi} (\rcd{\mu} A^{\mu})^2,
\end{equation}
where an additional gauge fixing term is added to fix the gauge redundancy in the Lorenz gauge~\cite{Motohashi}. 
The Euler-Lagrange equations then become~\cite{Birrell}
\begin{equation}
    \begin{aligned}
0 &= \rcd{\mu} F^{\mu \nu} + \frac{1}{\xi} \rcdu{\nu} \left( \rcd{\mu} A^{\mu} \right) \\
&= \square A^{\nu} - \rcd{\mu} \rcdu{\nu} A^{\mu} + \frac{1}{\xi} \rcdu{\nu} \left( \rcd{\mu} A^{\mu} \right),
    \end{aligned}
\end{equation}
where $\square = \rcd{\mu} \rcdu{\mu}$ is the d'Alembertian~\cite{Misner,Carroll}.
Using the commutation relation for torsion-free metrics~\cite{Misner,Carroll},
\begin{equation}
    \left[ \rcd{\mu}, \rcd{\nu} \right] A^{\rho} = R^{\rho}_{\phantom{\rho} \sigma \mu \nu} A^{\sigma},
\end{equation}
with the Riemann curvature tensor,
\begin{equation}
    R^{\rho}_{\phantom{\rho} \sigma \mu \nu} = \rpd{\mu} \Gamma^{\rho}_{\nu \sigma} - \rpd{\nu} \Gamma^{\rho}_{\mu \sigma} + \Gamma^{\rho}_{\mu \lambda} \Gamma^{\lambda}_{\nu \sigma} - \Gamma^{\rho}_{\nu \lambda} \Gamma^{\lambda}_{\mu \sigma},
\end{equation}
we can rewrite the wave equation as
\begin{equation}
    \square A^{\nu} - \left( 1 - \frac{1}{\xi} \right) \rcdu{\nu} \left( \rcd{\mu} A^{\mu} \right) - R_{\mu}^{\phantom{\mu} \nu} A^{\mu} = 0,
\end{equation}
where $R_{\mu \nu} = R_{(\mu \nu)} = R^{\lambda}_{\phantom{\lambda} \mu \lambda \nu}$ is the Ricci curvature tensor. 
Choosing the Fermi-Feynman gauge parameter $\xi = 1$~\cite{Tong_QFT,Birrell} and noting that $R_{\mu \nu} = 0$ for flat spacetime in any coordinate system (since $R^{\rho}_{\phantom{\rho} \sigma \mu \nu} = 0$~\cite{Misner}), the wave equation becomes
\begin{equation}
    \square A^{\mu} = 0.
\end{equation}
We can then lower the index of the four-potential using $\rcd{\alpha} g_{\mu \nu} = 0$~\cite{Misner,Carroll},
\begin{equation} \label{squareA}
    \square A_{\mu} = 0,
\end{equation}
such that we can identify $A_{\mu} = \left( \varphi, - \mathbf{A} \right)$ to be consistent with Secs.~\ref{Sec:DiracEqCurvedST} and~\ref{Sec:FieldHamiltonian}.

Inside the ring cavity, the modes must satisfy periodic boundary conditions such that the allowed frequencies when $\Omega = 0$ are given by
\begin{equation} \label{omega_n}
    \omega_n^I = \abs{\mathbf{k}_n}= \frac{2 \pi \abs{n}}{L_{\mathrm{path}}},
\end{equation}
where $n \in \{ \pm 1, \pm 2, \ldots \}$ and $L_{\mathrm{path}}$ is the path length of the cavity ($L_{\mathrm{path}} = 4 L$ for the square cavity of Fig.~\ref{Schematic}).
To quantize the E\&M field, we can then expand the potential in terms of creation and annihilation operators at a specific frequency and polarization as~\cite{Birrell,Tong_QFT,Greiner}
\begin{equation}
    \hat{A}_{\mu} = i \sum_{\mathbf{k}_n} \sum_{\zeta = 0}^3 \epsilon_{\mu}^{\zeta} (\mathbf{r}, \mathbf{k}_n) \left[ f_{\mathbf{k}_n}(\mathbf{r},t) \hat{a}_{\mathbf{k}_n, \zeta} - f_{\mathbf{k}_n}^* (\mathbf{r},t) \hat{a}_{\mathbf{k}_n, \zeta}^{\dagger} \right],
\end{equation}
where $\epsilon_{\mu}^{\zeta}$ is the polarization four-vector for different polarizations $\zeta$.
Note that for uniform rotation, the Bogoliubov transformation between inertial and rotating modes is trivial, so the same creation and annihilation operators may be used as in the inertial frame~\cite{Letaw2}. 
The polarizations can be taken to be~\cite{Tong_QFT,Greiner} one timelike, two transverse, and one longitudinal polarization which we denote with $\zeta = 0$, $\zeta = 1,2$, and $\zeta = 3$, respectively. 
The timelike and longitudinal polarizations are canceled out for physical states in the Lorenz gauge via the Gupta-Bleuler condition~\cite{Gupta,Bleuler,Blasone}.
The wave equation can then be written as
\begin{equation} \label{WaveEq_Pol}
    \epsilon_{\mu}^{\zeta} (\mathbf{k}_n) \square f_{\mathbf{k}_n} + 2 \left[ \rcdu{\nu} \epsilon_{\mu}^{\zeta} (\mathbf{k}_n) \right] \left[ \rpd{\nu} f_{\mathbf{k}_n} \right] + \left[ \square \epsilon_{\mu}^{\zeta} (\mathbf{k}_n) \right] f_{\mathbf{k}_n} = 0,
\end{equation}
where we have used $\rcd{\mu} f_{\mathbf{k}_n} = \rpd{\mu} f_{\mathbf{k}_n}$ for any scalar~\cite{Carroll}. 
We now note that the first term leads to Sagnac frequency shifts, as shown below, while the last two terms lead to polarization-transport effects in the form of spin- and helicity-rotation coupling~\cite{Mashhoon2,Mashhoon3,Hauck,Mashhoon4,Mashhoon5,Brodutch,Brodutch2,Zawistowski,Ruiz,Dahal,Mashhoon6,Mashhoon7,Mashhoon8,Mashhoon9}, similar to the spin-rotation coupling found for Dirac fields in Eq.~\eqref{GB_SpinConnection}.
Since the spin- and helicity-rotation coupling effects are generally smaller than the Sagnac effect by a factor of $\lambda / L_{\mathrm{path}}$ for optical light in the tetrad basis introduced in Sec.~\ref{Sec:GenBornMetric}~\cite{Mashhoon2,Mashhoon6,Dahal}, we now assume that the polarization vectors are constant in the local tetrad frame $\rcdu{\hat{\nu}} \epsilon^{\zeta}_{\hat{\mu}} (\mathbf{k}_n) \approx 0$ to leading order.
Upon making this assumption, the mode functions $f_{\mathbf{k}_n}$ must satisfy the wave equation~\cite{Carroll}
\begin{equation}
    \square f_{\mathbf{k}_n} = \frac{1}{\sqrt{- g}} \rpd{\mu} [\sqrt{- g} g^{\mu \nu} \rpd{\nu} f_{\mathbf{k}_n}] = 0,
\end{equation}
where we have again used $\rcdu{\mu} f_{\mathbf{k}_n} = \rpdu{\mu} f_{\mathbf{k}_n}$. 

We now specify the metric to be the generalized Born metric from Eq.~\eqref{g_GB}.
Using the ADM form from Eq.~\eqref{GenBorn_ADM}, the wave equation becomes
\begin{equation}
    \begin{aligned}
\square f_{\mathbf{k}_n} = 0 = & \rpd{t}^2 f_{\mathbf{k}_n} - \gamma^{ij} \rpd{i} \rpd{j} f_{\mathbf{k}_n} + 2 N^j \rpd{j} \rpd{t} f_{\mathbf{k}_n} \\
& + N^i N^j \rpd{i} \rpd{j} f_{\mathbf{k}_n} + (\rpd{j} N^i) N^j \rpd{i} f_{\mathbf{k}_n},
    \end{aligned}
\end{equation}
where we have used $\rpd{\mu} g^{00} = \rpd{0} g^{0j} = \rpd{0} g^{ij} = 0$ as well as $\rpd{j} N^j = 0$ for a constant rotation.
Then using 
\begin{equation}
    N^i \rpd{i} (N^j \rpd{j}) f_{\mathbf{k}_n} = N^i N^j \rpd{i} \rpd{j} f_{\mathbf{k}_n} + N^i (\rpd{i} N^j) \rpd{j} f_{\mathbf{k}_n},
\end{equation}
with $[\rpd{j}, N^j] = 0$, we find
\begin{equation}
    \begin{aligned}
0 &= \rpd{t}^2 f_{\mathbf{k}_n} - \bm{\nabla}^2 f_{\mathbf{k}_n} + 2 N^j \rpd{j} \rpd{t} f_{\mathbf{k}_n} +  N^i \rpd{i} (N^j \rpd{j}) f_{\mathbf{k}_n} \\
&= \left[ \rpd{t}^2 - \bm{\nabla}^2 - 2 \left( \left[\mathbf{\Omega} \cross \mathbf{r}\right] \bm{\cdot} \bm{\nabla} \right) \rpd{t} +  \left( \left[ \mathbf{\Omega} \cross \mathbf{r}\right] \bm{\cdot} \bm{\nabla} \right)^2 \right] f_{\mathbf{k}_n} \\
&= \left( \rpd{t} - \left[ \mathbf{\Omega} \cross \mathbf{r}\right] \bm{\cdot} \bm{\nabla} \right)^2 f_{\mathbf{k}_n} - \bm{\nabla}^2 f_{\mathbf{k}_n}.
    \end{aligned}
\end{equation}
Finally, we use a scalar triple product identity~\cite{Jackson} to rewrite this as
\begin{equation} \label{GenWaveEq}
    \square f_{\mathbf{k}_n} = \left( \rpd{t} - \mathbf{\Omega} \bm{\cdot} \left[ \mathbf{r} \cross \bm{\nabla} \right] \right)^2 f_{\mathbf{k}_n} - \bm{\nabla}^2 f_{\mathbf{k}_n} = 0.
\end{equation}
This can be compared to the wave equation for Minkowski spacetime where
\begin{equation} \label{NonRotWaveEq}
    \square \bar{f}_{\mathbf{k}_n} = \rpd{T}^2 \bar{f}_{\mathbf{k}_n} - \bm{\nabla}'^2 \bar{f}_{\mathbf{k}_n} = 0,
\end{equation}
and so $\rpd{T}^2 \longrightarrow \left( \rpd{t} - \mathbf{\Omega} \bm{\cdot} \left[ \mathbf{r} \cross \bm{\nabla} \right] \right)^2$ in the rotating frame. 
This modification has the form of a dot product of the angular frequency of the rotation with an angular momentum operator, and so can be identified as the Sagnac frequency shift for light in the rotating frame.

\subsection{Separation of variables} \label{Sec:SepVariables}
We now use the fact that the spacetime is stationary such that it has a timelike Killing vector inside the light cylinder [see Eq.~\eqref{KillingMag}].
This, in turn, means that a Lie derivative of the metric tensor along the Killing vector field vanishes~\cite{Carroll}, so that the metric components are independent of the associated time coordinate. 
Consequently, the field equations are invariant under time translations, and so the mode functions may be separated into temporal and spatial parts~\cite{Birrell,Wald},
\begin{equation}
    f_{\mathbf{k}_n} = \mathcal{N}_{\mathbf{k}_n} e^{- i \omega_n t} u_{\mathbf{k}_n} (\mathbf{r}),
\end{equation}
where $\mathcal{N}_{\mathbf{k}_n}$ is a normalization factor~\cite{Steck}.
For a general rotation axis and ring cavity geometry, we cannot use separation of variables on the spatial mode function $u_{\mathbf{k}_n}$ along the plane of the cavity. 
Nevertheless, the cavity still defines a closed optical loop and so counterpropagating modes will still experience opposite Sagnac shifts proportional to the projection of $\mathbf{\Omega}$ along the area vector of the ring cavity~\cite{Post,Wang,Ori}. 

To simplify the mode functions further, we now make three simplifying assumptions (see Fig.~\ref{Schematic}):
\begin{enumerate}[label=\roman*.]
    \item the rotation axis is normal to the plane of the ring cavity, $\mathbf{\Omega} \approx \tilde{\Omega} \mathbf{z} = \Omega \cos \theta \mathbf{z}$;

    \item the ring cavity is approximately a circular cavity of radius $\mathcal{R}$;

    \item and the field is confined in the radial and $z$ directions such that the ring cavity can be treated as a ring of radius $\mathcal{R}$.
\end{enumerate}
With assumption i, the generalized Born metric Eq.~\eqref{g_GB} reduces to the typical Born metric Eq.~\eqref{g_B}.
Therefore, the wave equation Eq.~\eqref{GenWaveEq} reduces to 
\begin{equation}
    \left( \rpd{t} - \tilde{\Omega} \rpd{\phi} \right)^2 f_{\mathbf{k}_n} - \frac{1}{\rho^2} \rpd{\phi}^2 f_{\mathbf{k}_n} - \rpd{\rho}^2 f_{\mathbf{k}_n} - \rpd{z}^2 f_{\mathbf{k}_n} - \frac{1}{\rho} \rpd{\rho} f_{\mathbf{k}_n} = 0.
\end{equation}
The Born metric is axisymmetric $\rpd{\phi} g^{\mathrm{B}}_{\mu \nu} = 0$ and symmetric in the $z$-direction $\rpd{z} g^{\mathrm{B}}_{\mu \nu} = 0$, and so there are spacelike Killing vectors associated with rotations in $\phi$ and translations in $z$~\cite{Carroll}. 
Therefore, we can use the same argument as the timelike Killing vector above, as well as assumption ii, to separate the mode function as~\cite{Konoplya,Brill,Letaw}
\begin{equation}
    f_{\mathbf{k}_n} = \tilde{\mathcal{N}}_{\mathbf{k}_n} e^{- i \omega_n t} e^{i m_{\mathbf{k}_n} \phi} \mathfrak{R}_{m_{\mathbf{k}_n}} (\rho) e^{i k_z z},
\end{equation}
where $\tilde{\mathcal{N}}_{\mathbf{k}_n} = \mathcal{N}_{\mathbf{k}_n} / \sqrt{\mathcal{V}}$ includes the quantization volume $\mathcal{V}$ such that the spatial mode functions are normalized.
Lastly, we use assumption iii to ignore the radial and $z$ components of the mode function such that
\begin{equation} \label{ModeFunc}
    f_{\mathbf{k}_n} = \tilde{\mathcal{N}}_{\mathbf{k}_n} e^{- i \omega_n t} e^{i m_{\mathbf{k}_n} \phi},
\end{equation}
while the path length in Eq.~\eqref{omega_n} becomes $L_{\mathrm{path}} = 2 \pi \mathcal{R}$.
Thus, the wave equation becomes
\begin{equation}
    \left( \rpd{t} - \tilde{\Omega} \rpd{\phi} \right)^2 f_{\mathbf{k}_n} - \frac{1}{\mathcal{R}^2} \rpd{\phi}^2 f_{\mathbf{k}_n} = 0,
\end{equation}
and so
\begin{equation}
    \omega_n^2 + 2 \tilde{\Omega} m_{\mathbf{k}_n} \omega_n - m_{\mathbf{k}_n}^2 \left( \frac{1}{\mathcal{R}^2} - \tilde{\Omega}^2 \right) = 0.
\end{equation}
We then obtain
\begin{equation}
    \omega_n = - \tilde{\Omega} m_{\mathbf{k}_n} \pm \frac{\abs{m_{\mathbf{k}_n}}}{\mathcal{R}},
\end{equation}
and since (in SI units) $c/\mathcal{R} > \tilde{\Omega}$, positive frequency modes satisfy
\begin{equation}
    \omega_n = \frac{\abs{m_{\mathbf{k}_n}}}{\mathcal{R}} - m_{\mathbf{k}_n} \tilde{\Omega}.
\end{equation}
If one uses assumptions ii and iii on the non-rotating wave equation Eq.~\eqref{NonRotWaveEq}, one finds the inertial frequency
\begin{equation}
    \omega_n^I = \frac{\abs{m_{\mathbf{k}_n}}}{\mathcal{R}},
\end{equation}
such that the rotating-frame frequencies are given by
\begin{equation}
    \omega_n = \omega_n^I - m_{\mathbf{k}_n} \tilde{\Omega}. 
\end{equation}
Next, we write the spatial mode function as $e^{i k s}$ for a coordinate of propagation $s$.
For photons propagating in the azimuthal direction, we have $s = \mathcal{R} \phi$ such that we can equate $m_{\mathbf{k}_n} = k_n \mathcal{R}$ in Eq.~\eqref{ModeFunc}.
Since the wave vector will be along the $\pm \phi$ directions, co-rotating photons ($+$) will have $k_n = \abs{{\mathbf{k}_n}}$ while counter-rotating photons ($-$) will have $k_n = - \abs{{\mathbf{k}_n}}$, and so
\begin{equation}
    \omega_{n,\pm} = \omega_n^I \mp \abs{k_n} \tilde{\Omega} \mathcal{R},
\end{equation}
which demonstrates the opposite Sagnac shift for counterpropagating modes. 
Finally, the mode function is given by
\begin{equation} \label{SagnacModeFunc}
    f_{\mathbf{k}_n} = \tilde{\mathcal{N}}_{\mathbf{k}_n} \exp \left[ - i \left( \omega_n^I - k_n \tilde{\Omega} \mathcal{R} \right) t \right] \exp \left[ i k_n \mathcal{R} \phi \right],
\end{equation}
such that the quantized field becomes
\begin{equation}
    \begin{aligned}
& \hat{A}_{\mu} = \sum_n \sum_{\zeta = 0}^3 \epsilon_{\mu}^{\zeta} (\mathbf{k}_n) \tilde{\mathcal{N}}_{\mathbf{k}_n} \left[ i e^{- i \left( \omega_n^I - k_n \tilde{\Omega} \mathcal{R} \right) t} e^{i k_n \mathcal{R} \phi} \hat{a}_{\mathbf{k}_n, \zeta} + \mathrm{H.c.} \right] \\
&= \sum_{n > 0} \sum_{\zeta = 0}^3 \left( \epsilon_{\mu}^{\zeta} (\mathbf{k}_n) \tilde{\mathcal{N}}_{\mathbf{k}_n} \left[ i e^{- i \left( \omega_n^I - \abs{k_n} \tilde{\Omega} \mathcal{R} \right) t} e^{i \abs{k_n} \mathcal{R} \phi} \hat{a}_{\mathbf{k}_n, \zeta} + \mathrm{H.c.} \right] \right. \\
& + \left. \epsilon_{\mu}^{\zeta} (\mathbf{k}_{- n}) \tilde{\mathcal{N}}_{\mathbf{k}_{- n}} \left[ i e^{- i \left( \omega_n^I + \abs{k_n} \tilde{\Omega} \mathcal{R} \right) t} e^{- i \abs{k_n} \mathcal{R} \phi} \hat{a}_{\mathbf{k}_{- n}, \zeta} + \mathrm{H.c.} \right] \right).
    \end{aligned}
\end{equation}

\subsection{Gauge transformation}
We now transform from the Lorenz gauge to the Coulomb gauge to then use in the PZW transformation in Sec.~\ref{Sec:JaynesCummingsHydrogen}. 
The gauge transformation is performed according to~\cite{Griffiths_EM}
\begin{equation}
    A_{\mu}^{\mathrm{L}} \longrightarrow A_{\mu}^{\mathrm{C}} = A_{\mu}^{\mathrm{L}} + \partial_{\mu} \chi,
\end{equation}
with the gauge function $\chi$ chosen so that the transformed vector potential satisfies the Coulomb condition $\bm{\nabla} \bm{\cdot} \mathbf{A}_{\mathrm{C}} = 0$.
This gauge transformation is done explicitly in Refs.~\cite{Jackson2,Wundt}, so we only present the main results here. 
Notably, the gauge transformation eliminates the longitudinal polarization in the vector potential~\cite{Wundt} such that $\mathbf{A}_{\mathrm{C}} = \mathbf{A}_{\mathrm{C}, \perp}$.
Meanwhile, the timelike polarization is no longer treated as a propagating mode and is instead absorbed into the instantaneous scalar potential $\varphi_{\mathrm{C}}$ in the non-covariant Coulomb gauge~\cite{Weinberg2,Greiner,Haller}. 
For an empty cavity in the absence of atoms, the scalar potential is a solution to Laplace's equation which we can then set to zero $\varphi_{\mathrm{C}} = 0$.
Therefore, the ring cavity's field may be entirely represented by the transverse components of the vector potential,
\begin{equation} \label{A_tot_app}
    \begin{aligned}
\hat{\mathbf{A}} = \sum_n \sum_{\zeta = 1}^2 \bm{\epsilon}_{\zeta} (\mathbf{k}_n) \tilde{\mathcal{N}}_{\mathbf{k}_n} \left[ i e^{- i \left( \omega_n^I - k_n \tilde{\Omega} \mathcal{R} \right) t} e^{i k_n \mathcal{R} \phi} \hat{a}_{\mathbf{k}_n, \zeta} + \mathrm{H.c.} \right],
    \end{aligned}
\end{equation}
where we have dropped the ``$\mathrm{C}$'' label as we will now work entirely in the Coulomb gauge.

We can now obtain the normalization factors on the mode functions by enforcing the Coulomb gauge commutation relation~\cite{Tong_QFT}
\begin{equation} \label{A_Pi_comm}
    \begin{aligned}
\left[ \hat{A}_{\mu} (\mathbf{r}), \hat{\Pi}^{\nu} (\mathbf{r}') \right] &= \sqrt{\gamma} i \hbar \delta_{\mu}^j \delta_k^{\nu} \left( \delta_j^k - \frac{\partial_j \partial^k}{\partial_i \partial^i} \right) \delta^{(3)}(\mathbf{r} - \mathbf{r}') \\
&= \sqrt{\gamma} i \hbar \delta_{\mu}^j \delta_k^{\nu} \delta_{jk}^{\perp}(\mathbf{r} - \mathbf{r}'), 
    \end{aligned}
\end{equation}
such that we have the vector commutator
\begin{equation}
    \left[ \hat{\mathbf{A}} (\mathbf{r}), \hat{\bm{\Pi}} (\mathbf{r}') \right] = - \sqrt{\gamma} i \hbar \delta^{\perp}(\mathbf{r} - \mathbf{r}').
\end{equation}
Here, the Dirac delta functions are coordinate-dependent which is why the $\sqrt{\gamma}$ appears, playing the role of a Jacobian factor in $\int d^3r \sqrt{\gamma} \delta^{(3)} (\mathbf{r} - \mathbf{r'}) f(\mathbf{r}) = f(\mathbf{r}')$.
The conjugate momentum is obtained from Eq.~\eqref{ConjMom} 
\begin{equation}
    \begin{aligned}
\hat{\bm{\Pi}} = & \sqrt{\gamma} \left[ - \bm{\nabla} \hat{\varphi} - \rpd{t} \hat{\mathbf{A}} + \mathbf{N} \cross \left( \bm{\nabla} \cross \hat{\mathbf{A}} \right) \right] \\
= & - \sqrt{\gamma} \sum_n \sum_{\zeta = 1}^2 \omega_n^I \bm{\epsilon}_{\zeta} (\mathbf{k}_n) \tilde{\mathcal{N}}_{\mathbf{k}_n} \cross \\
& \left[ e^{- i \left( \omega_n^I - k_n \tilde{\Omega} \mathcal{R} \right) t} e^{i k_n \mathcal{R} \phi} \hat{a}_{\mathbf{k}_n, \zeta} + \mathrm{H.c.} \right],
    \end{aligned}
\end{equation}
which gives (in SI units)
\begin{equation} \label{NormFactor}
    \mathcal{N}_{\mathbf{k}_n} = \sqrt{\frac{\hbar}{2 \epsilon_0 \omega_n^I}},
\end{equation}
where we have approximated the Dirac delta function as a sum, $\delta^{(3)}(\mathbf{r} - \mathbf{r}') \sim \sum_{\mathbf{k}_n} \exp[i \mathbf{k}_n \bm{\cdot} (\mathbf{r} - \mathbf{r}')] / \mathcal{V}$, for discretized modes.
Note that we then have $\mathcal{N}_{\mathbf{k}_n} = \mathcal{N}_{\mathbf{k}_{- n}}$.

\subsection{Two-mode approximation}
Finally, we assume that there is only one relevant inertial frequency, labeled $\omega_c$, in the atom-field interaction of interest, i.e., one frequency is near resonance while all other frequencies are far-detuned: $\abs{\omega_a - \omega_c} \ll \abs{\omega_a - \omega_n^I}, \; \forall \omega_n^I \neq \omega_c$. 
We can then drop every mode of the cavity except the modes with frequencies $\omega_{\pm} = \omega_c \mp \abs{k} \tilde{\Omega} \mathcal{R}$ in the co-rotating ($\omega_+$) and counter-rotating ($\omega_-$) directions, where we have dropped the subscript on the wave vector amplitude. 
We then assume only one polarization in each direction, labeled $\epsilon_{\pm}$, is relevant to the atom-field interaction.
With these approximations, the vector potential in the Coulomb gauge can be simplified to
\begin{equation} \label{A_tm_app}
    \begin{aligned}
\hat{\mathbf{A}} &= \tilde{\mathcal{N}} \bm{\epsilon}_+ \left[ i e^{- i \left( \omega_c - \abs{k} \tilde{\Omega} \mathcal{R} \right) t} e^{i \abs{k} \mathcal{R} \phi} \hat{a}_+ + \mathrm{H.c.} \right] \\
& + \tilde{\mathcal{N}} \bm{\epsilon_-} \left[ i e^{- i \left( \omega_c + \abs{k} \tilde{\Omega} \mathcal{R} \right) t} e^{- i \abs{k} \mathcal{R} \phi} \hat{a}_- + \mathrm{H.c.} \right],
    \end{aligned}
\end{equation}
where $\tilde{\mathcal{N}} \equiv \tilde{\mathcal{N}}_k = \tilde{\mathcal{N}}_{- k}$ (and $\mathcal{N} \equiv \sqrt{\mathcal{V}} \tilde{\mathcal{N}}$).

We can now solve for the field Hamiltonian derived in Sec.~\ref{Sec:FieldHamiltonian},
\begin{equation}
    \hat{H}_{\mathrm{F}} = \int_{\mathcal{V}} d^3 r \sqrt{\gamma} \normalorder{\left( \frac{1}{2} \left[ \hat{\mathbf{E}}^2 + \hat{\mathbf{B}}^2 \right] - \mathbf{\Omega} \bm{\cdot} \left[ \mathbf{r} \cross \left( \hat{\mathbf{E}} \cross \hat{\mathbf{B}} \right) \right] \right)},
\end{equation}
where we have applied normal ordering $\normalorder{\hat{a}_{\pm} \hat{a}_{\pm}^{\dagger}} = \hat{a}_{\pm}^{\dagger} \hat{a}_{\pm}$ and $\normalorder{\hat{a}_{\pm}^{\dagger} \hat{a}_{\pm}} = \hat{a}_{\pm}^{\dagger} \hat{a}_{\pm}$ which neglects the vacuum energy. 
First, we find the magnetic field,
\begin{equation} \label{B_quant}
    \begin{aligned}
\hat{\mathbf{B}} &= \bm{\nabla} \cross \hat{\mathbf{A}} \\
&= - \abs{k} \tilde{\mathcal{N}} (\bm{\phi} \cross \bm{\epsilon}_+) \left[ e^{- i \left( \omega_c - \abs{k} \tilde{\Omega} \mathcal{R} \right) t} e^{i \abs{k} \mathcal{R} \phi} \hat{a}_+ + \mathrm{H.c.} \right] \\
& + \abs{k} \tilde{\mathcal{N}} (\bm{\phi} \cross \bm{\epsilon}_-) \left[ e^{- i \left( \omega_c + \abs{k} \tilde{\Omega} \mathcal{R} \right) t} e^{- i \abs{k} \mathcal{R} \phi} \hat{a}_- + \mathrm{H.c.} \right],
    \end{aligned}
\end{equation}
where we have used the vector calculus identity~\cite{Jackson} $\bm{\nabla} \cross (C \mathbf{a}) = (\bm{\nabla} C) \cross \mathbf{a} + C (\bm{\nabla} \cross \mathbf{a})$ and again assumed the polarization vectors are approximately constant in the local tetred frame such that $\bm{\nabla} \cross \bm{\epsilon}_{\pm} \approx 0$.
We then obtain the electric field via the conjugate momentum,
\begin{equation} \label{E_quant}
    \begin{aligned}
\hat{\mathbf{E}} = & \frac{\hat{\bm{\Pi}}}{\sqrt{\gamma}} = - \rpd{t} \hat{\mathbf{A}} - \left( \mathbf{\Omega} \cross \mathbf{r} \right) \cross \hat{\mathbf{B}} \\
= & - \omega_c \tilde{\mathcal{N}} \bm{\epsilon}_+ \left[ e^{- i \left( \omega_c - \abs{k} \tilde{\Omega} \mathcal{R} \right) t} e^{i \abs{k} \mathcal{R} \phi} \hat{a}_+ + \mathrm{H.c.} \right] \\
& - \omega_c \tilde{\mathcal{N}} \bm{\epsilon_-} \left[ e^{- i \left( \omega_c + \abs{k} \tilde{\Omega} \mathcal{R} \right) t} e^{- i \abs{k} \mathcal{R} \phi} \hat{a}_- + \mathrm{H.c.} \right],
    \end{aligned}
\end{equation}
where we have used $\bm{\phi} \bm{\cdot} \bm{\epsilon}_{\pm} = 0$ for transverse modes and again set $\hat{\varphi} = 0$ for the empty cavity.
From these (and using the discretized Dirac delta function), we obtain
\begin{equation}
    \frac{1}{2} \int_{\mathcal{V}} d^3 r \sqrt{\gamma} \normalorder{\left( \hat{\mathbf{E}}^2 + \hat{\mathbf{B}}^2 \right)} = 2 \omega_c^2 \mathcal{N}^2 \left( \hat{a}_+^{\dagger} \hat{a}_+ + \hat{a}_-^{\dagger} \hat{a}_- \right),
\end{equation}
while the angular momentum from Eq.~\eqref{ell},
\begin{equation}
    \begin{aligned}
\hat{\bm{\ell}} = & \omega_c \abs{k} \tilde{\mathcal{N}}^2 [\mathbf{r} \cross \bm{\phi}] \left( 2 \hat{a}_+^{\dagger} \hat{a}_+ - 2 \hat{a}_-^{\dagger} \hat{a}_- \right. \\
& + \left[ e^{- 2 i \left( \omega_c - \abs{k} \tilde{\Omega} \mathcal{R} \right) t} e^{2 i \abs{k} \mathcal{R} \phi} \hat{a}_+^2 + \mathrm{H.c.} \right] \\
& \left. - \left[ e^{- 2 i \left( \omega_c + \abs{k} \tilde{\Omega} \mathcal{R} \right) t} e^{- 2 i \abs{k} \mathcal{R} \phi} \hat{a}_-^2 + \mathrm{H.c.} \right] \right),
    \end{aligned}
\end{equation}
gives the Sagnac frequency shifts
\begin{equation}
    \int_{\mathcal{V}} d^3 r \normalorder{\mathbf{\Omega} \bm{\cdot} \bm{\hat{\ell}}} = 2 \omega_c \abs{k} \tilde{\Omega} \mathcal{R} \mathcal{N}^2 \left( \hat{a}_+^{\dagger} \hat{a}_+ - \hat{a}_-^{\dagger} \hat{a}_- \right).
\end{equation}
Plugging in the normalization factors from Eq.~\eqref{NormFactor}, these then give the total field Hamiltonian (in SI units)
\begin{equation} \label{H_F_app}
    \hat{H}_{\mathrm{F}} = \hbar \left( \omega_c - \abs{k} \tilde{\Omega} \mathcal{R} \right) \hat{a}_+^{\dagger} \hat{a}_+ + \hbar \left( \omega_c + \abs{k} \tilde{\Omega} \mathcal{R} \right) \hat{a}_-^{\dagger} \hat{a}_-.
\end{equation}
We also plug the normalization factor into the electric field to find (again in SI units)
\begin{equation} \label{E_quant_final}
    \begin{aligned}
\hat{\mathbf{E}} = & - \sqrt{\frac{\hbar \omega_c}{2 \epsilon_0 \mathcal{V}}} \bm{\epsilon}_+ \left[ e^{- i \left( \omega_c - \abs{k} \tilde{\Omega} \mathcal{R} \right) t} e^{i \abs{k} \mathcal{R} \phi} \hat{a}_+ + \mathrm{H.c.} \right] \\
& - \sqrt{\frac{\hbar \omega_c}{2 \epsilon_0 \mathcal{V}}} \bm{\epsilon_-} \left[ e^{- i \left( \omega_c + \abs{k} \tilde{\Omega} \mathcal{R} \right) t} e^{- i \abs{k} \mathcal{R} \phi} \hat{a}_- + \mathrm{H.c.} \right],
    \end{aligned}
\end{equation}
which can then be inserted into the electric dipole interaction in Sec.~\ref{Sec:JaynesCummingsHydrogen}.

\bibliography{references.bib}

\end{document}